\documentclass{emulateapj}
\usepackage{natbib,times}
\usepackage{lscape}

\slugcomment{Received 2009 January 26; accepted 2009 June 4; published 2009 July 9 on {\bf ApJS 183, 197-213. Further modifications on Table 1 and 2.}}

\shorttitle{}
\shortauthors{}
\begin{document}

\title{Galaxy clusters identified from the SDSS DR6 and their properties}

\author{Z. L. Wen\altaffilmark{1,2}, 
        J. L. Han\altaffilmark{1}
        and 
        F. S. Liu\altaffilmark{3}}

\altaffiltext{1}{National Astronomical Observatories, Chinese Academy of Sciences, 
                 20A Datun Road, Chaoyang District, Beijing 100012, People's Republic of China; 
                 zhonglue@bao.ac.cn, hjl@bao.ac.cn.} 
\altaffiltext{2}{Graduate University of the Chinese Academy of Sciences, 
                 Beijing, 100049, People's Republic of China}
\altaffiltext{3}{College of Physics Science and Technology, 
                 Shenyang Normal University, Shenyang 110034, People's Republic of China; 
                 lfs@bao.ac.cn.}

%%%%%%%%%%%%%%%%%%%%%%%%%%%%%%%%%%%%%%%%%%%%%%%%%%%%%%%%%%%%%%%%%%%%%%%%%%%%%

\begin{abstract}

Clusters of galaxies in most previous catalogs have redshifts
$z\le0.3$. Using the photometric redshifts of galaxies from the Sloan
Digital Sky Survey Data Release 6 (SDSS DR6), we identify 39,668
clusters in the redshift range $0.05< z <0.6$ with more than eight
luminous ($M_r\le-21$) member galaxies.
Cluster redshifts are estimated accurately with an uncertainty less
than 0.022. The contamination rate of member galaxies is found to be
roughly 20\%, and the completeness of member galaxy detection reaches
to $\sim$90\%. Monte Carlo simulations show that the cluster detection
rate is more than 90\% for massive ($M_{200}>2\times10^{14}~M_{\odot}$)
clusters of $z\le0.42$. The false detection rate is $\sim$5\%.
We obtain the richness, the summed luminosity, and the gross galaxy
number within the determined radius for identified clusters. They are
tightly related to the X-ray luminosity and temperature of
clusters. Cluster mass is related to the richness and summed
luminosity with $M_{200}\propto R^{1.90\pm0.04}$ and $M_{200}\propto
L_r^{1.64\pm0.03}$, respectively. In addition, 685 new candidates of
X-ray clusters are found by cross-identification of our clusters with
the source list of the {\it ROSAT} X-ray survey.

\end{abstract}


%%%%%%%%%%%%%%%%%%%%%%%%%%%%%%%%%%%%%%%%%%%%%%%%%%%%%%%%%%%%%%%%%%%%%%%
\section{Introduction}

As the largest gravitational bound systems in the universe, clusters
of galaxies are important tracers to study the large scale structure
\citep{bah88,phg92,cye+96,bfc97}. Statistical studies of clusters
constrain the cosmological parameters, for example, $\Omega_m$, the mass
density parameter of the universe, and $\sigma_8$, the amplitude of
mass fluctuations at a scale of 8 $h^{-1}$ Mpc
\citep{rb02,sel02,dah06,pd07,rdn07}.  The detailed studies of clusters
provide the strong evidence of dark matter and constrain the abundance
of dark matter in the universe
\citep[see, e.g.,][]{ief+96,cs03,jfi+07,bse+08}. Clusters are also
important laboratories to investigate the evolution of galaxies in
dense environment, e.g., the Butcher--Oemler effect, the
morphology--density relation
\citep{dre80,bo78,bo84,gma99,goy+03,gyf+03}. In addition, clusters can
act as efficient gravitational lenses and provide an independent way
to study high-redshift faint background galaxies
\citep[see, e.g.,][]{bki+99,sib+02,mkm+03,sek+04}.

A lot of clusters have been found in various surveys in last decades.
By visual inspection of optical images, \citet{abe58} was the first to
identify a large sample of rich clusters from the National
Geographic Society--Palomar Observatory Sky Survey. The catalog was
improved and expanded to 4073 rich clusters by \citet{aco89}. Some
other catalogs of clusters were obtained visually from optical
images \citep[see, e.g.,][]{zhw68,gho86}.

To reduce subjectivity, an automated peak-finding method was developed
by \citet{she85} and applied to the Edinberg/Durham survey
\citep{lnc+92} and the Automatic Plate Measurement Facility survey
\citep{dms97}. A matched-filter algorithm was later developed by
\citet{plg+96} and applied to the Palomar Distant Cluster Survey, and
later the Edinburgh/Durham Southern Galaxy Catalogue \citep{bnp00},
the Sloan Digital Sky Survey (SDSS) data \citep{kkp+02}, and the
Canada--France--Hawaii Telescope Legacy Survey
\citep{obc+07}. \citet{gcl+03} used an adaptive kernel technique
\citep{sil86} to search for clusters in the galaxy sample
($15.0<m_r<19.5$) of the digitized Second Palomar Observatory Sky
Survey and presented the NSC catalog containing 8155 clusters of
$z\le 0.3$ in the sky region of 5800 deg$^2$. \citet{ldg+04}
incorporated the adaptive kernel and the Voronoi tessellation
techniques \citep{rbf+01,kkp+02} to a deeper sample ($m_r<21.1$) of
the digitized Second Palomar Observatory Sky Survey and presented the
NSCS catalog containing 9956 clusters of $0.1\le z\le 0.5$ in the sky
region of $2700$ deg$^2$.

The above methods were applied to detect clusters in single-band
imaging data, and suffered severe contamination from foreground and
background galaxies.  To reduce projection effect, several methods
have been developed to search for clusters in multicolor photometric
data and have been successfully used to the Red-Sequence Cluster
Survey \citep{gy00,gy05} and the SDSS \citep{gsn+02,mnr+05,kma+07a}.

When spectroscopic redshifts are available for a large sample of
galaxies, clusters or groups can be identified in three dimensions
conventionally by the friend-of-friend algorithm
\citep{hg82,gh83}. Many catalogs of clusters or groups have been
obtained from the various redshift surveys: \citet{tul87} for the
Nearby Galaxies Catalog, \citet{rzz+99} for the ESO Slice Project,
\citet{toh+00} for the Las Campanas Redshift Survey, \citet{gmc+00}
for the Nearby Optical Galaxy Sample, \citet{rgp+02} for the Southern
Sky Redshift Survey, \citet{mz02}, \citet{ebc+04}, and \citet{ymv+05}
for the two-degree field Galaxy Redshift Survey (2dfGRS),
\citet{gnd+05} for the DEEP2 Galaxy Redshift Survey, and \citet{mz05},
\citet{bfw+06}, \citet{ymv+07}, \citet{dhj+07}, and \citet{tes+08} for
the SDSS. A matched-filter algorithm was developed in spectroscopic or 
photometric redshift surveys \citep{wk02} and applied to the Two Micron 
All Sky Survey (2MASS) data \citep{kwh+03}.

The SDSS \citep{yaa+00} offers an opportunity to produce the largest
and most complete cluster catalog. It provides photometry in five broad
bands ($u$, $g$, $r$, $i$, and $z$) covering 10,000 deg$^2$ and
the follow-up spectroscopic observations. The photometric data reach
a limit of $r=22.5$ \citep{slb+02} with the star--galaxy separation
reliable to a limit of $r=21.5$ \citep{lgi+01}. The spectroscopic
survey observes galaxies with an extinction-corrected Petrosian
magnitude of $r<17.77$ for the main galaxy sample \citep{swl+02} and
$r<19.5$ for the Luminous Red Galaxy (LRG) sample \citep{eag+01}.  The
spectroscopic data of the SDSS enable to detect clusters up to
$z\sim0.1$, while the photometric data enable to detect clusters up to
$z\sim0.5$ \citep{bma+03}.

\citet{mz05} performed the friend-of-friend algorithm to the
spectroscopic data of the SDSS DR3 and obtained 10,864 groups with a
richness (i.e., the number of member galaxies) $\ge4$. Similarly,
\citet{bfw+06} obtained three volume-limited samples from the SDSS
DR3, which contain 4107, 2684, and 1357 groups with a richness $\ge3$
out to redshift of 0.1, 0.068, and 0.045, respectively. The catalogs
by \citet{dhj+07} and \citet{tes+08} contain 11,163 groups with a
richness $\ge4$ and 50,362 groups with a richness $\ge2$. Using a
modified friend-of-friend algorithm by \citet{ymv+05}, \citet{wby+06}
identified 53,229 groups of $z\le0.2$ with a mass greater than
$3\times10^{11}~h^{-1}~M_{\odot}$ from the SDSS DR2, and later
\citet{ymv+07} obtained 301,237 groups of $z\le0.2$ with a mass
greater than $6.3\times10^{11}~h^{-1}~M_{\odot}$ from the SDSS DR4. By
using merely spectroscopic data of the SDSS, most of the groups in
\citet{wby+06} and \citet{ymv+07} have only one member galaxy.

Searching for galaxies in seven-dimensional position and color spaces,
\citet{mnr+05} presented the C4 catalog, which contains 748 clusters
of $z\le 0.12$ with a richness $\ge10$ from the spectroscopic data of
the SDSS DR2. To reduce incompleteness due to the SDSS spectroscopic
selection bias, e.g., fiber collisions, \citet{yss+08} incorporated
the spectroscopic and photometric data to search for density peaks 
and obtained 924 clusters from the SDSS DR5 in the redshift range
$0.05<z<0.1$. The SDSS photometric data give a large space for cluster
finding. From the photometric data of the SDSS Early Data Release
(SDSS EDR), \citet{gsn+02} used the ``Cut and Enhance'' method to
detect the enhanced densities for galaxies of similar colors and
obtained 4638 clusters of $z<0.4$. \citet{kkp+02} developed a hybrid
matched-filter cluster finder and applied it to the SDSS EDR. The detected
clusters were compiled by \citet{bma+03}. By looking for small and
isolated concentrations of galaxies, \citet{lat+04} identified 175
compact groups with a richness between 4 and 10 from the SDSS
EDR. \citet{kma+07a} developed a ``Red-Sequence cluster finder'', the
maxBCG, to detect clusters dominated by red galaxies. From the SDSS
DR5, \citet{kma+07b} obtained a complete volume-limited catalog
containing 13,823 clusters in the redshift range
$0.1<z<0.3$. Recently, \citet{dpg+08} presented a modified adaptive
matched-filter algorithm to identify clusters, which is adaptive to
imaging surveys with spectroscopic redshifts, photometric redshifts, and
no redshift information at all. Tests of the algorithm on the mock
SDSS catalogs suggest that the detected sample is $\sim$85\% complete
for clusters with masses above $1.0\times10^{14}~M_{\odot}$ up to
$z=0.45$.

Most of the clusters in above catalogs have been identified in optical
bands at $z\le 0.3$. For methods based on the single-band image data,
clusters at higher redshifts are difficult to detect due to projection
effect. In multicolor surveys, the color cut is an efficient method to
detect clusters since the red sequence, i.e., the color--magnitude
relation, can be used as an indicator of redshift. For example,
\citet{kma+07a} used the $g-r$ color cut to detect clusters in the
SDSS data. At $0.1<z<0.3$, the $g-r$ color difference is sensitive to
redshift because of the shift of the 4000 \AA\ break between $g$ and
$r$ bands. However, the 4000 \AA\ break migrates into the $r$ band at
$z>0.35$, then the $g-r$ color difference is insensitive to redshift.

Galaxy clusters can be detected by other approaches. The X-ray
observation is an efficient and independent way to identify 
clusters with a low contamination rate
\citep[see, e.g.,][]{sch78,ghm+90,eeb+98}. About 1100 X-ray clusters have
been identified from the {\it ROSAT} survey, including the Northern ROSAT
All-Sky cluster sample \citep[NORAS;][]{bvh+00}, the ROSAT-ESO flux
limit cluster sample \citep[REFLEX;][]{bsg+04} and the {\it ROSAT} PSPC 400
deg$^2$ cluster sample \citep{bvh+07}. From a sample of 495 {\it ROSAT}
X-ray extended sources, \citet{bvh+00} presented the NORAS sample
containing 376 clusters with count rates of $C_X\ge 0.06$ count
s$^{-1}$ in the 0.1--2.4 keV band. The REFLEX is a complete sample,
containing 447 X-ray clusters in the southern hemisphere with a flux
limit of $3\times 10^{-12}$~erg~s$^{-1}$ cm$^{-2}$ in the 0.1--2.4 keV
band \citep{bsg+04}. \citet{bvh+07} presented a catalog of X-ray
clusters detected in a new {\it ROSAT} PSPC survey. From $\sim$400 deg$^2$,
they identified 287 extended X-ray sources with a flux limit of
$1.4\times10^{-13}$~erg~s$^{-1}$ cm$^{-2}$ in the 0.5--2 keV band, of
which 266 are optically confirmed as galaxy clusters, groups or
elliptical galaxies. Besides the X-ray method, the Sunyaev--Zeldovich
effect and the weak lensing effect have been tried to search for clusters
\citep{sch96,cjg+00,wtm+01,pah+05}.

Usually, cluster richness is indicated by the number of cluster
member. The spectroscopic redshifts are required to accurately
determine the member galaxies of clusters. However, spectroscopic
redshifts are usually flux-limited. Only clusters at low redshifts
have their richnesses well determined
\citep[see, e.g.,][]{bfw+06,mnr+05}. Moreover, the fiber collision in the
SDSS sometimes results in the incompleteness of spectroscopic data
about 35\% or even worse for clusters of $z\le0.1$ \citep{yss+08}.
Without redshifts, the richnesses were generally measured by the
number of all galaxies in a projected radius for the clusters selected
from single-band image data, and hence suffered from heavy projection
effect. For example, the Abell richness is defined to be the number of
galaxies within 2 mag range below the third-brightest galaxy within a
radius of 1.5~$h^{-1}$ Mpc \citep{aco89}. Without accurate member
discrimination, few cluster catalogs have richness well
determined.
In multicolor survey, it is possible to discriminate cluster galaxies
by color cuts with contamination partly being
excluded. \citet{kma+07b} discriminated member galaxies based on
cluster ridgeline for the SDSS maxBCG clusters. They defined the
richness to be the number of galaxies brighter than $0.4L^{\ast}$
within $\pm2\sigma_c$ of the ridgeline defined by the BCG color. Here
$\sigma_c$ is the error of the measured color.

For many researches, such as large-scale structure studies, a
volume-limited cluster sample with richness well determined in a broad
redshift range is required. The cluster-finding algorithm need to
maximize the completeness of member galaxies and minimize the
contamination from foreground and background galaxies. Previous
studies \citep{bl00,yzc+01,yzj03,zat+03,yzy+04,wyy+07} showed that
most of luminous member galaxies of clusters can be picked out using
photometric redshifts. In this paper, we identify clusters from the
SDSS photometric data by discriminating member galaxies in the
photometric redshift space. Our method is valid to the multicolor
surveys for which photometric redshifts can be estimated. Clusters can
be detected even up to $z\sim0.6$ in the SDSS.

This paper is organized as follows. In Section 2, we describe our
cluster-finding algorithm in the photometric redshift space. In Section 3,
we examine the statistical properties of our cluster catalog. Using the
SDSS spectroscopic data, we estimate the uncertainty of cluster
redshift, the contamination rate, and the completeness of discriminated
member galaxies of clusters. Monte Carlo simulations are performed to
estimate cluster detection rate and false detection rate of our
algorithm.  In Section 4, we compare our catalog with the previous
optical-selected cluster catalogs. In Section 5, we discuss the
correlations between the richness and summed luminosity of clusters
with the measurements in X-rays. New candidates of X-ray clusters are
extracted by the cross-identification of our clusters with the source list
in the {\it ROSAT} All Sky Survey. A summary is presented in Section 6.

Throughout this paper, we assume a $\Lambda$CDM cosmology, taking
$H_0=$100 $h$ ${\rm km~s}^{-1}$ ${\rm Mpc}^{-1}$, with $h=0.72$, 
$\Omega_m=0.3$ and $\Omega_{\Lambda}=0.7$.

%%%%%%%%%%%%%%%%%%%%%%%%%%%%%%%%%%%%%%%%%%%%%%%%%%%%%%%%%%%%%%%%%%%%%%%%%%%%
\section{The Cluster Detection}

In the traditional friend-of-friend algorithm, clusters and their
member galaxies are identified in spectroscopic redshift space with
appropriately chosen linking lengths both in line of sight and
perpendicular directions. However, spectroscopic redshift surveys are
usually flux-limited; thus, the detected cluster/group samples are
obtained from flux-limited galaxy samples. Complete volume-limited
samples can be obtained only at low redshifts by the SDSS
spectroscopic data \citep[see, e.g.,][]{bfw+06}. When spectroscopic
redshifts are not available for the faint galaxies, photometric
redshifts can be used. We now attempt to identify clusters using the
photometric redshift catalog of the SDSS DR6 in a broad redshift range
($z\sim0.05$--0.6).

\subsection{Photometric redshifts in the SDSS}

\begin{figure}
\epsscale{1.}
\plotone{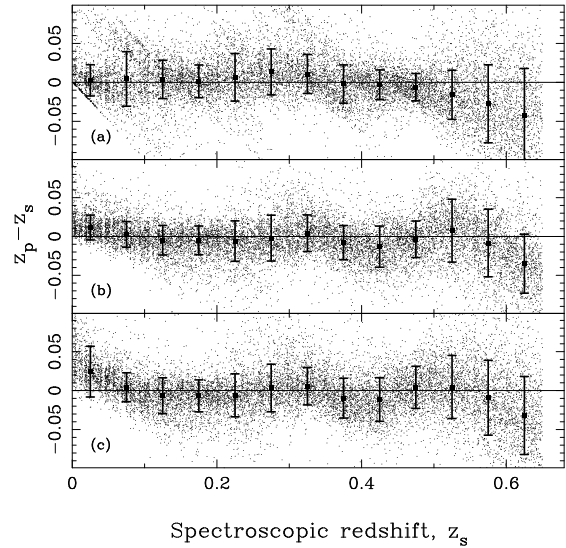}
\caption{Comparison between spectroscopic redshift $z_s$ with
  photometric redshift $z_p$. Panel (a) is for \citet[][version
    v1.6]{cbc+03}, and panels (b) and (c) are for CC2 and D1 from \citet{olc+08}.
\label{phzg}}
\end{figure}

Based on the SDSS photometric data, photometric redshifts of galaxies
brighter than $r=22$ have been estimated by two groups. \citet{cbc+03}
provided photometric redshifts utilizing various techniques, from
empirical to template and hybrid techniques. \citet{olc+08} estimated
photometric redshifts with the Artificial Neural Network technique 
and provided two different photometric redshift estimates, CC2 and D1.
Figure~\ref{phzg} shows the differences between photometric
and spectroscopic redshifts at $z<0.65$. The galaxy sample is selected
from the SDSS spectroscopic data at $z\le0.4$ and from the 2dF-SDSS
Luminous Red Galaxy Survey \citep{cde+06} at $z>0.4$.
The error bars show the uncertainties of photometric redshifts,
$\sigma_{68}$, the ranges containing 68\% sample in the distribution
of $|z_p-z_s|$. We find that the uncertainties for three
estimates are comparable, being 0.02--0.03 at $z<0.5$ and $\sim$0.07
at $z\ge 0.5$.
At $z\le0.3$, the estimate by \citet[][version v1.6, see panel a in
  Figure~\ref{phzg}]{cbc+03} has more photometric redshifts with large
deviations than the CC2 and D1 estimates. At $0.3<z<0.5$, the
scattering by \citet{cbc+03} is smaller than those of the CC2 and D1
estimates. For both CC2 and D1 estimates, photometric redshifts are
systematically larger than the spectroscopic redshifts at $z\sim0.3$
and 0.5 but smaller at $z\sim0.4$. In our cluster-finding algorithm,
the linearity between photometric and spectroscopic redshifts is
important. The systematic biases can induce systematic underestimation
or overestimation on the density of galaxies in the photometric
redshift space, thus affecting the uniformity of cluster selection. The
estimate by \citet{cbc+03} has smaller systematic derivation in
general except at $z>0.5$. To obtain an uniform cluster detection in a
broad redshift range, we adopt the photometric redshifts by
\citet{cbc+03} in the following cluster detection.

Most of the galaxies at $z>0.2$ in the SDSS spectroscopic data are the
luminous red galaxies \citep{eag+01}, which have strong continuum
feature, the 4000 \AA\ break. Because of this feature, photometric
redshifts are well estimated for these galaxies. However, there is no
sample of less luminous galaxies for the calibration of photometric
redshifts at $z>0.2$. The uncertainties of photometric redshifts
should be larger for less luminous galaxies of $z>0.2$ due to the
shallower depth of the 4000 \AA\ break and larger photometric
errors. In the following analysis, we assume that the uncertainty,
$\sigma_z$, of photometric redshift increases with redshift in the
form of $\sigma_z=\sigma_0(1+z)$ for all galaxies.

\subsection{Cluster finding algorithm}
\label{algorithm}

The galaxy sample is taken from the SDSS DR6 database, including the
coordinates (R.A., Decl.), the model magnitudes with $r\le21.5$, and
the photometric redshifts, the $K$-corrections and absolute magnitudes
estimated by \citet{cbc+03}.
To obtain a volume-limited cluster catalog, we consider only the
luminous galaxies of $M_r\le-21$. We assume that they are member
galaxy candidates of clusters. Our cluster-finding algorithm includes
the following steps:

1. For each galaxy at a given $z$, we assume that it is the central
galaxy of a cluster candidate, and count the number of luminous
``member galaxies'' of $M_r\le-21$ within a radius of 0.5 Mpc and a
photometric redshift gap between $z\pm0.04(1+z)$.
Within this redshift gap, most of the member galaxies of a cluster can
be selected, with a completeness of $\sim$80\% if assuming the
photometric redshift uncertainty of $\sigma_z=0.03(1+z)$.
The radius of 0.5 Mpc is chosen to get a high overdensity level and a
low false detection rate according to simulation tests (see
Section~\ref{falsrate}). It is smaller than the typical radius of a
rich cluster, but a rich cluster can have enough luminous member
galaxies within this radius for detection.

2. To avoid a cluster identified repeatedly, we consider only one
cluster candidate within a radius of 1 Mpc and a redshift gap of
0.1. We define the center of a cluster candidate to be the
position of the galaxy with a maximum number count. If two or more
galaxies show the same maximum number counts, we take the brightest
one as the central galaxy. The cluster redshift is defined to be the
median value of the photometric redshifts of the recognized
``members''.

3. For each cluster candidate at $z$, all galaxies within 1 Mpc from the
cluster center and the photometric redshift gap between
$z\pm0.04(1+z)$ are assumed to be the member galaxies, and then their
absolute magnitudes are recalculated with the cluster redshift.

\begin{figure}
\epsscale{1.}
\plotone{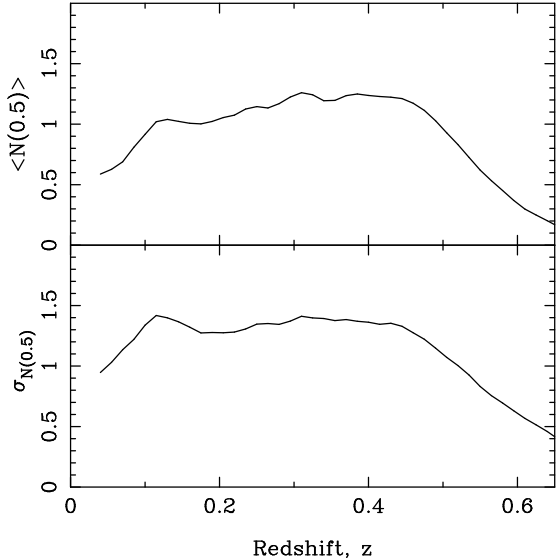}
\caption{Mean and rms of the number counts as a function of
  redshift for background galaxies of $M_r\le-21$ within a radius of
  0.5 Mpc and a redshift gap between $z\pm0.04(1+z)$.
\label{flat}}
\end{figure}

\begin{figure}
\epsscale{1.}
\plotone{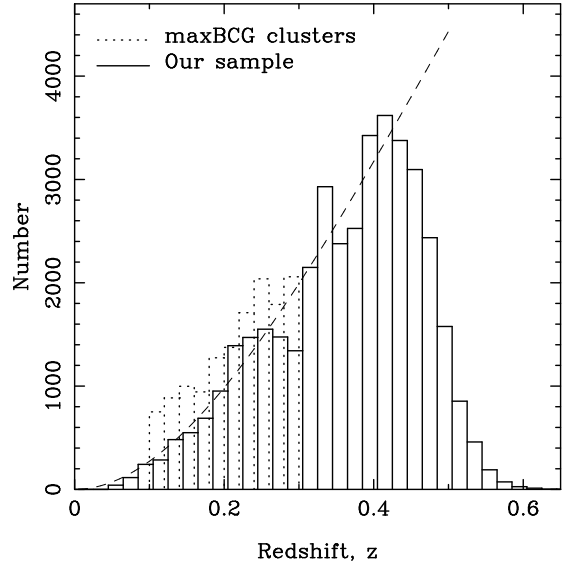}
\caption{Redshift distribution of clusters in our catalog. The dotted
  histogram in the redshift range $0.1<z<0.3$ is for the SDSS maxBCG
  clusters. The dashed line is the expected distribution for a
  complete volume-limited sample.
\label{hist}}
\end{figure}

\begin{figure}
\epsscale{1.15}
\plotone{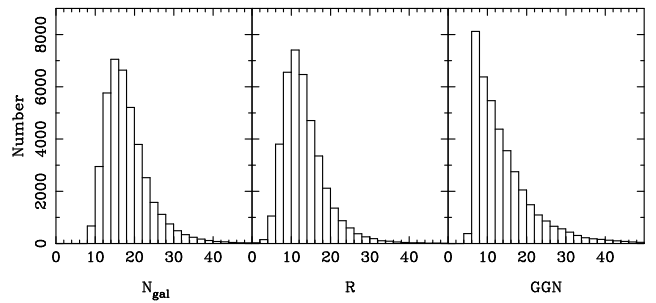}
\caption{Distributions of the number of member galaxy candidates
  within a radius of 1 Mpc ({\it $N_{\rm gal}$}, left), the cluster
  richness ({\it $R$}, middle) and the gross galaxy number ({\it
    $GGN$, right}) for clusters in our catalog.
\label{hisrich}}
\end{figure}

\begin{figure}
\epsscale{1.2}
\plotone{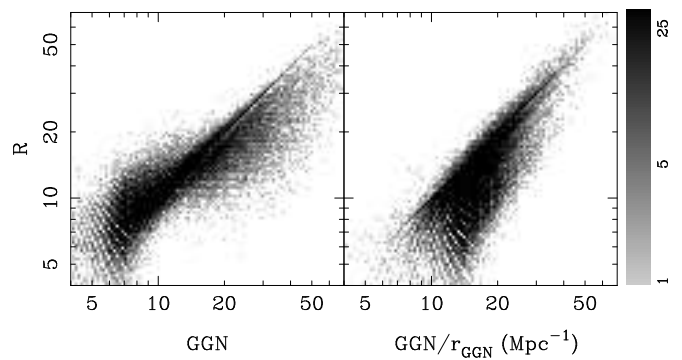}
\caption{Comparison between the gross galaxy number ({\it $GGN$,
    left}) and $GGN/r_{\rm GGN}$ ({\it right}) with cluster
  richness. The density of cluster sample is indicated by grey in the
  plot.
\label{ggn}}
\end{figure}

\begin{figure*}
\begin{center}
\resizebox{76mm}{!}{\includegraphics{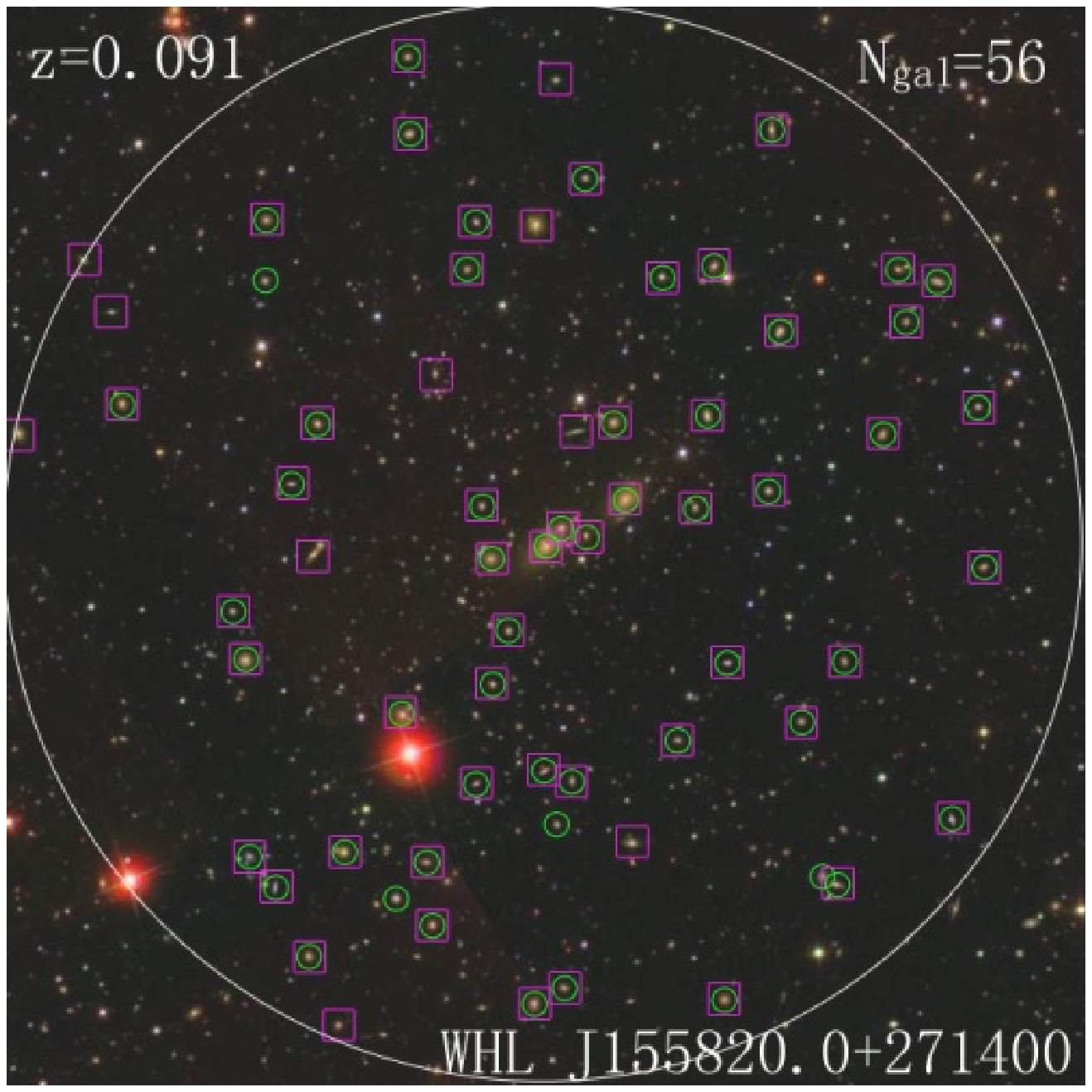}}
\resizebox{76mm}{!}{\includegraphics{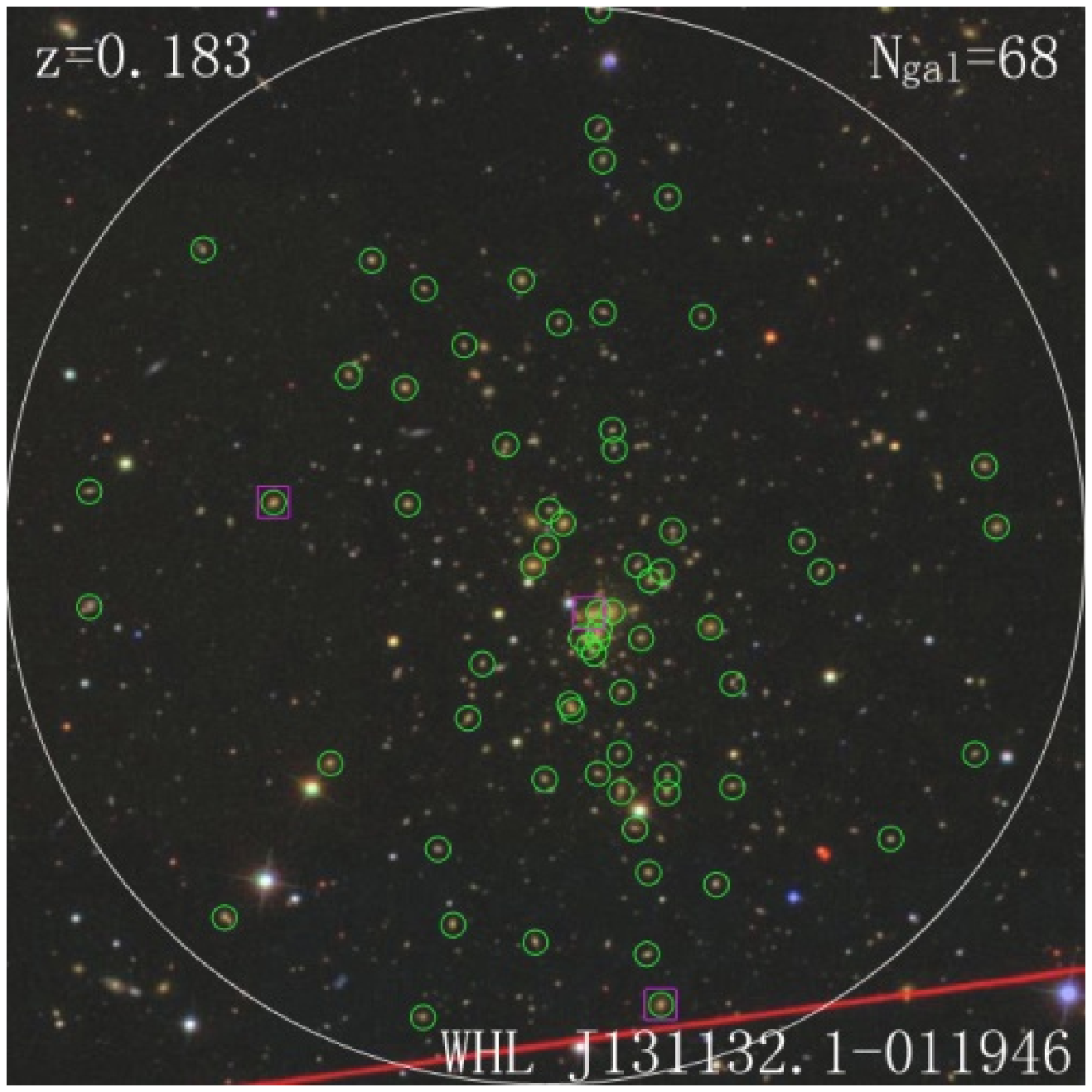}}\\[1mm]
\resizebox{76mm}{!}{\includegraphics{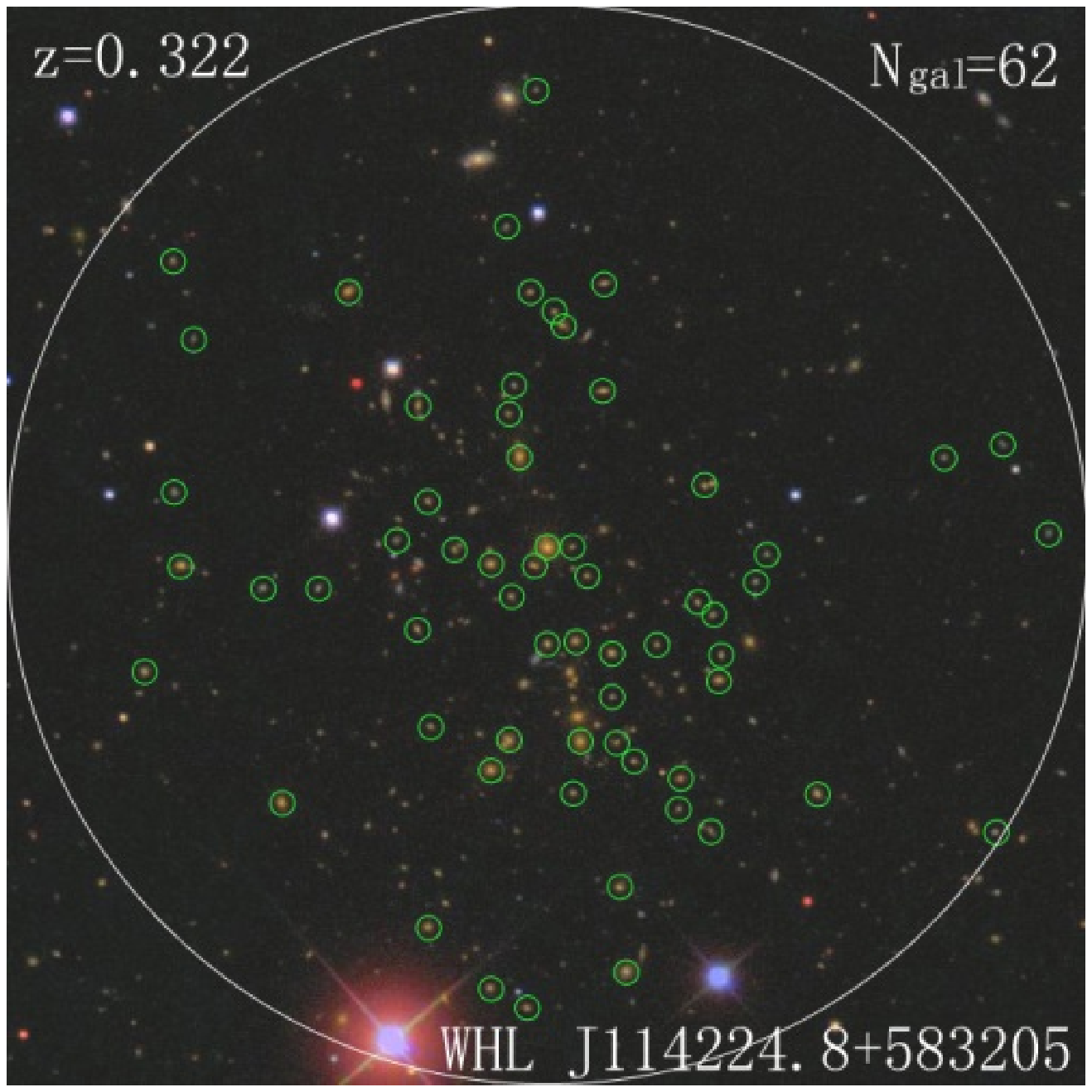}}
\resizebox{76mm}{!}{\includegraphics{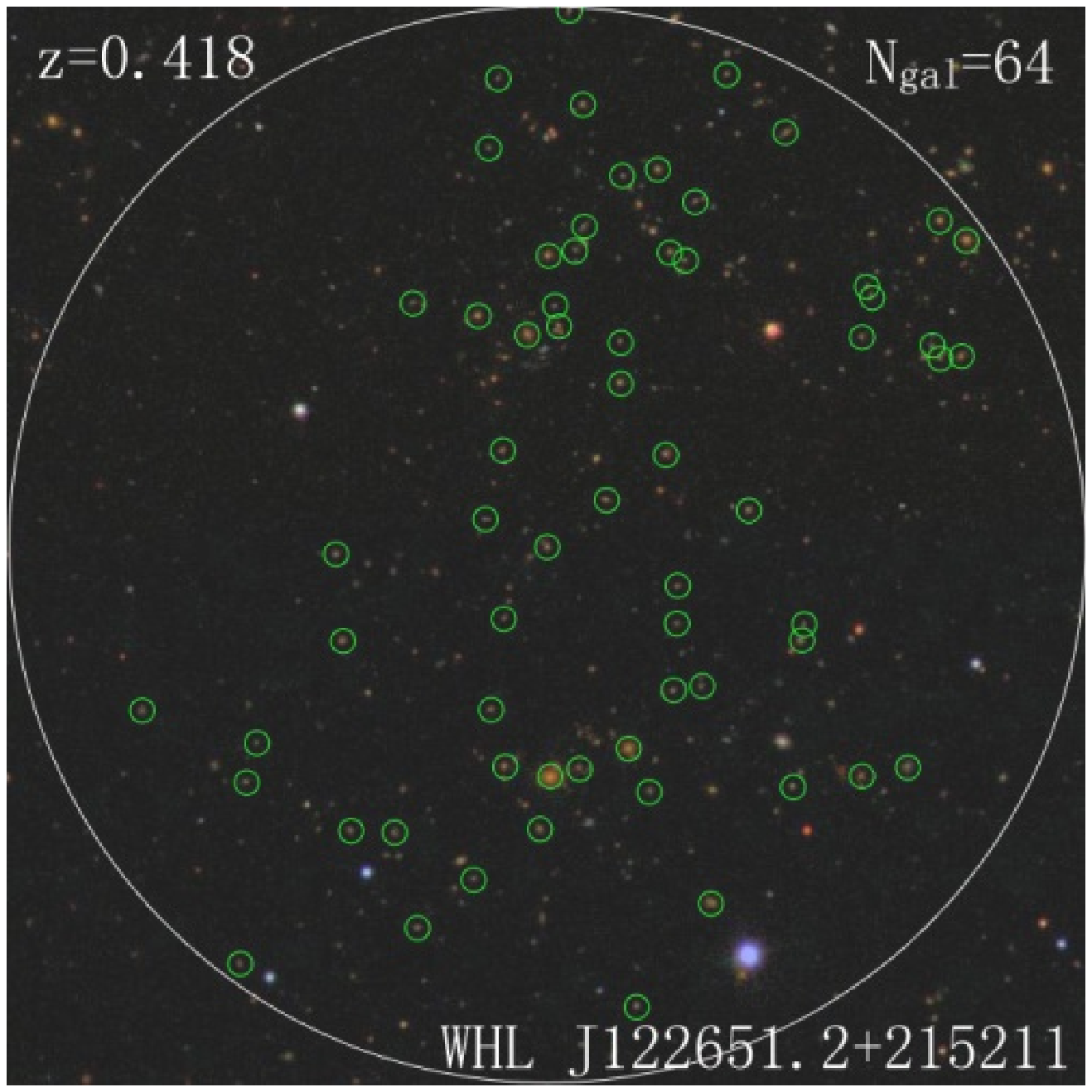}}\\[1mm]
\resizebox{76mm}{!}{\includegraphics{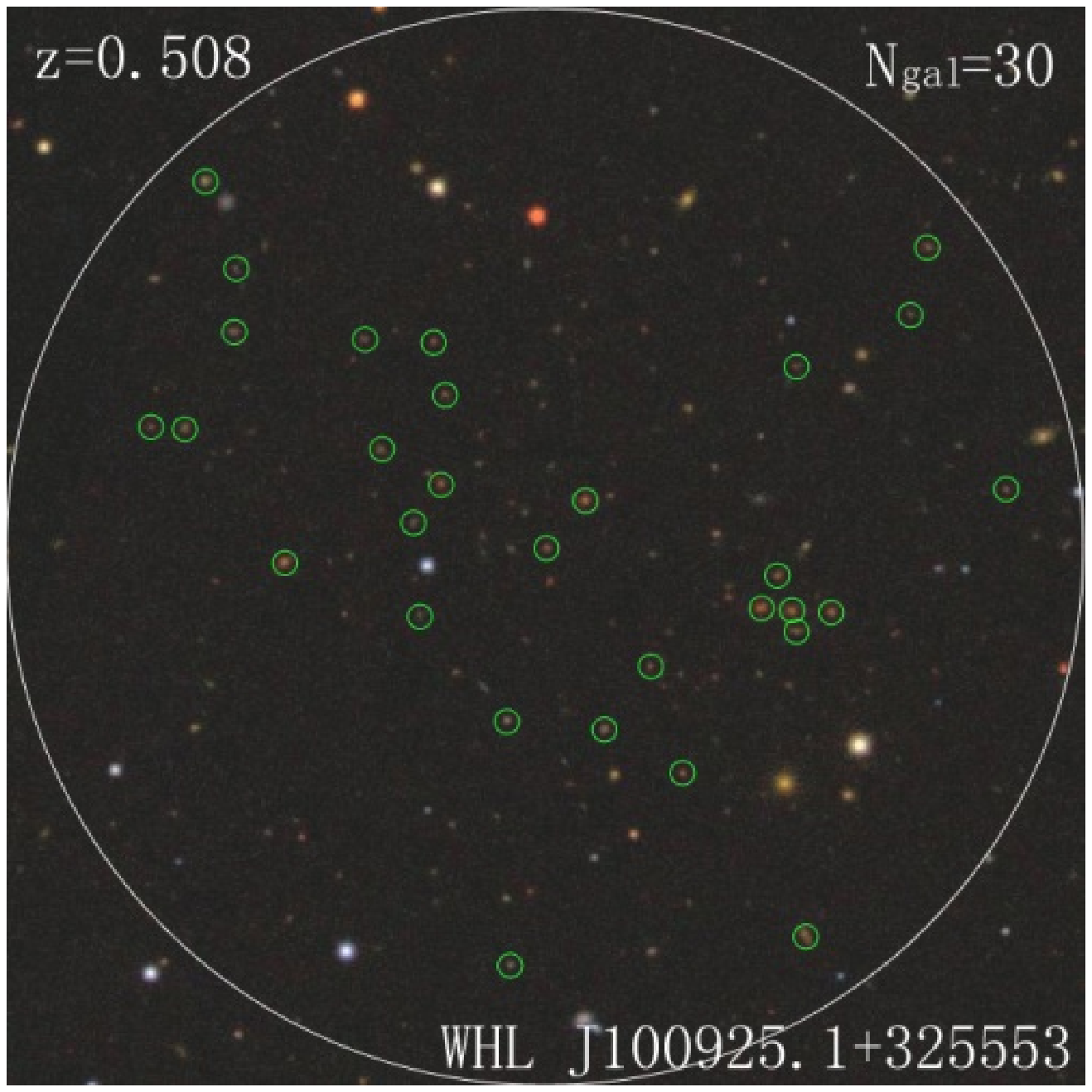}}
\resizebox{76mm}{!}{\includegraphics{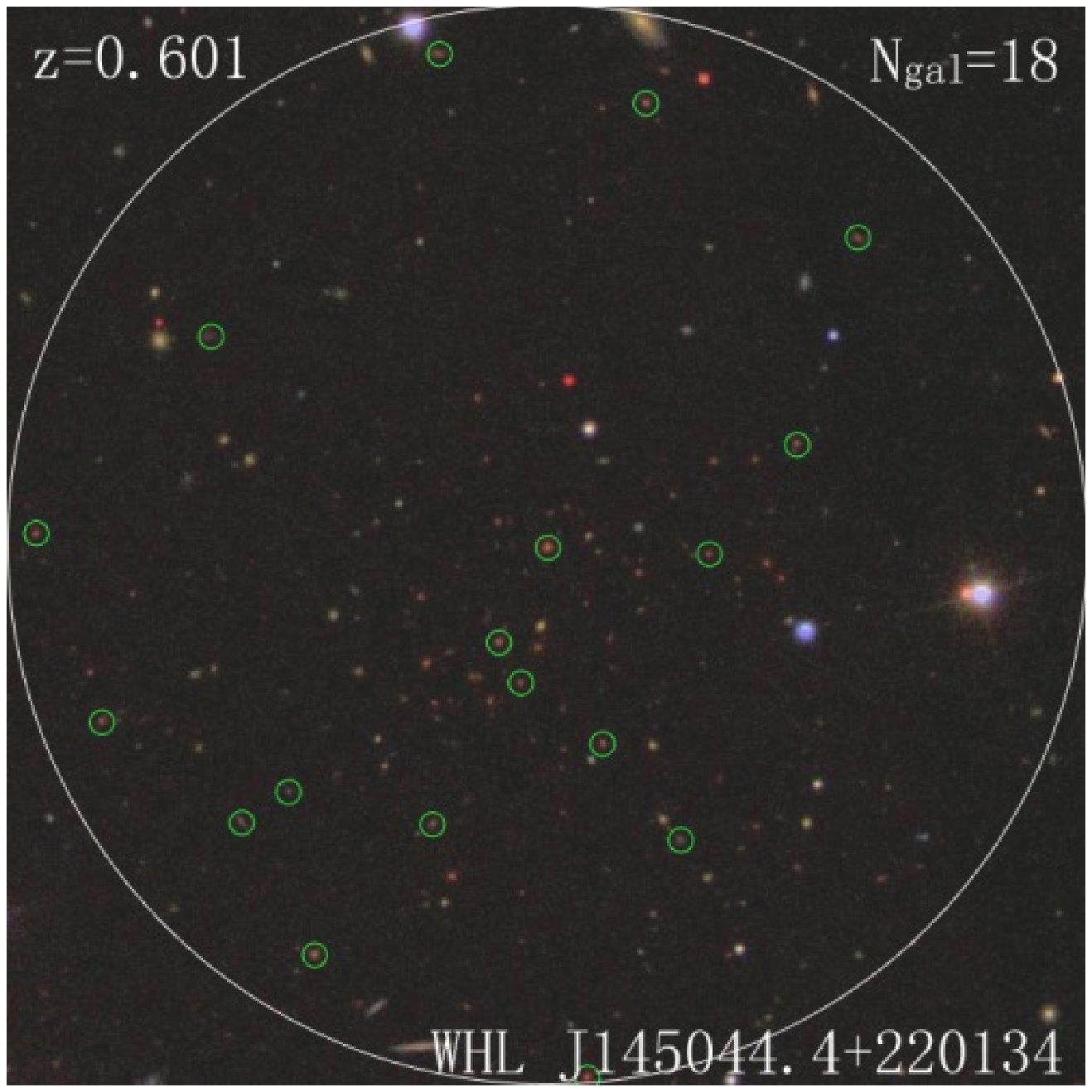}}\\[1mm]
\caption{Examples of detected clusters at different redshifts and
  their member galaxies discrimination. The big circle on the image
  has a radius of 1 Mpc from the cluster center. The small circles
  indicate member candidates discriminated by our method. The squares
  indicate member galaxies of $M_r\leq-21$ with velocities differing
  from that of the cluster by less than 4500~km~s$^{-1}$ in the SDSS
  spectroscopic data. {\it A color version of this figure is available
    in the online journal.}
\label{example}}
\end{center}
\end{figure*}

We detect a cluster if more than eight member galaxies of $M_r\le-21$ are
found within 0.5 Mpc from the cluster center. For clusters
at very low redshifts, although most of the member galaxies of $M_r\le
-21$ can be included within the photometric redshift gap, their
absolute magnitudes have large uncertainties when the estimated
redshift slightly deviates from its true redshift. Therefore, we
restrict our cluster detection with a lower redshift cutoff of
$z=0.05$. The nearby clusters ($z<0.05$) have been easily detected in
the spectroscopic redshift space \citep[see, e.g.,][]{mnr+05}.

To show overdensity of our clusters, we estimate the mean number
counts of galaxies within the same criteria of our algorithm, and the
root mean square (rms). At a given $z$, 2000 random positions (R.A.,
Decl.) are selected in the real background of galaxies. We count the
number of galaxies ($M_r\le-21$), $N(0.5)$, within a radius of 0.5 Mpc
and a redshift gap between $z\pm0.04(1+z)$, and then estimate the mean
number count and the rms (see Figure~\ref{flat}).
The mean number counts, $\langle N(0.5)\rangle$, is found to be
$\sim$1.2 in the redshift range $0.1<z<0.42$, and the rms is also
nearly constant, to be $\sigma_{N(0.5)}\sim1.3$ at $0.1<z<0.42$.  The
number counts decrease at higher redshift ($z>0.42$) because the
galaxy sample with a faint end of $r=21.5$ is incomplete for galaxies
of $M_r\le-21$.  We define the overdensity level of a cluster to
be $D=(N(0.5)-\langle N(0.5)\rangle)/\sigma_{N(0.5)}$. The minimum
number is eight within 0.5 Mpc from a cluster center, and corresponds to 
a minimum overdensity level $D$ about 4.5, above which the false
detection rate is very low in principle (see Section~\ref{falsrate}).

The number of member galaxy candidates ($N_{\rm gal}$ hereafter) is
defined to be the number of galaxies ($M_r\le-21$) within 1 Mpc (not 0.5
Mpc) from the cluster center in the redshift gap between
$z\pm0.04(1+z)$. The cluster richness, $R$, is defined by the number
of real member galaxies in this region. It is estimated by $N_{\rm
  gal}$ but subtracting contamination from foreground and background
galaxies. The contamination has to be estimated according to the local
background for each cluster. First, for each cluster, we divide the
area from its center to a radius of 3 Mpc into 36 annuluses, each with
an equal area of 0.25$\pi$ Mpc$^2$, and then count the number of
luminous ($M_r\le-21$) galaxies within each annulus. Certainly, this is
done within the redshift gap $z\pm0.04(1+z)$. Secondly, we get the
distribution peak at the count $n$ from the 36 number counts. The
background is estimated from the average galaxy density in all
annuluses with a number count less than $n+\sigma_{N(0.5)}\approx
n+1$. More galaxies in an annulus are probably from real structures
around the cluster, such as merging clusters, superclusters or
cosmological web structures. The average contamination background
within an annulus area of 0.25$\pi$ Mpc$^2$ is
\begin{equation}
\langle N_{\rm cb}\rangle=\Big[\sum_{i=1}^{36} N^i_{\rm ann}\theta(n+1-N^i_{\rm ann})\Big]\Big/N_{\rm ring}.
\end{equation}
Here $\theta(x)$ is the step function, $\theta(x)=1$ for $x\ge0$ and zero
otherwise; $N^i_{\rm ann}$ is the number count within $i$th
annulus; $N_{\rm ring}$ is the {\it total number of annuluses} with
$N^i_{\rm ann}\le n+1$. Then, the real number of cluster galaxies
(richness, $R$) within a radius of 1 Mpc is estimated to be
$R=N_{\rm gal}-4\times\langle N_{\rm cb}\rangle$.

We notice that for many clusters, the radius of 1 Mpc is not the
boundary of luminous cluster galaxies. The boundary can be recognized
from the number counts within the annuluses. It is defined to
be the radius of the first annulus from a cluster center, from which
two outer successive annuluses have $N^i_{\rm ann}\le n+1$.  We take
it as the radius for member galaxy detection, $r_{\rm GGN}$, within
which we count all luminous galaxies. After subtracting background, we
obtain the gross galaxy number of a cluster, $GGN$.

From the SDSS DR6, we obtain 39,668 clusters (named after WHL and J2000
coordinates of cluster center) in the redshift range $0.05<z<0.6$. All
clusters are listed in Table~\ref{cat} (a full list is available in
the online version). Figure~\ref{hist} shows the redshift distribution
of the clusters, compared with that of the SDSS maxBCG clusters. The
distribution can be well fitted by the expected distribution for a
complete volume-limited sample (the dashed line) with a number density of
$7.8\times 10^{-6}$~Mpc$^{-3}$ at $z<0.42$. Above this redshift, it is
less complete because of the flux cutoff at $r=21.5$ for the input
galaxy sample.

Figure~\ref{hisrich} shows the distributions of the number of member
galaxy candidates within a radius of 1 Mpc, $N_{\rm gal}$, the cluster
richness, $R$, and the gross galaxy number, $GGN$. The peaks are
at $N_{\rm gal}\sim$16, $R\sim$10 and $GGN\sim6$. Among 39,668 clusters
listed in our catalog, 28,082 clusters (71\%) have a richness $R\ge10$,
4059 clusters (10\%) have $R\ge20$, and 610 clusters (1.5\%) have
$R\ge30$.

Figure~\ref{ggn} compares $GGN$ and $GGN/r_{\rm GGN}$ with cluster
richness. We find that $GGN$ is related to cluster richness but not
linearly, while $GGN/r_{\rm GGN}$ is nearly linearly related to
cluster richness. The scatter is larger at the lower end probably
because of the quantized radius of annuluses, which is more uncertain
at smaller radius. We notice that cluster--galaxy cross-correlation is
described by a power law, $\xi (r)\propto r^{-\gamma}$, with the
correlation index $\gamma\sim2$ \citep[e.g.,][]{le88}. Hence, the
value of $GGN/r_{\rm GGN}$ is related to the amplitude of
cluster--galaxy cross-correlation, which has been shown to be a tracer
of cluster richness \citep{yl99}. In the following, we use the
richness $R$ to study the statistical properties of our catalog and
compare them with other optical catalogs, but we will consider $GGN$
and $GGN/r_{\rm GGN}$ in the discussions of their correlations with
X-ray properties (see Section~\ref{opt_xray}).

Examples of six clusters at different redshifts and their member
galaxies discrimination are shown in Figure~\ref{example}. For the
cluster WHL J155820.0$+$271400 (Abell 2142) at $z=0.091$, we get 
$N_{\rm gal}=56$ and $R=44$. Within a radius of 1 Mpc, 62 
galaxies of $M_r\leq-21$ have velocities differing from that of the
cluster by less than 4500~km~s$^{-1}$ in the SDSS spectroscopic data
(the velocity dispersion of a very rich cluster can be
1500~km~s$^{-1}$). We discriminate 52 (84\%) of them by using photometric
redshifts. In addition, four galaxies with velocity difference greater
than 4500~km~s$^{-1}$ are selected as members. This example shows that
photometric redshifts are reliable for member galaxies discrimination
at $z\sim0.1$.
The richness of this cluster is 95 by the maxBCG method (defined to be
the number of member galaxies brighter than 0.4$L^{\ast}$ within 1
$h^{-1}$ Mpc), and 164 by the method of \citet{ymv+07} (defined to be
the number of member galaxies of $M_r\le-19.5$).
The second example is WHL J131132.1$-$011946 (Abell 1689) at
$z=0.183$, for which we get $N_{\rm gal}=68$ and $R=62.74$. Only three member
galaxies have spectroscopic redshifts in the SDSS. The richness of
this cluster is 102 by the maxBCG method, but only two by the method of
\citet{ymv+07}.
For clusters WHL J114224.8$+$583205 (Abell 1351) at $z=0.322$, WHL
J122651.2$+$215211 (NSCS) at $z=0.418$ and WHL J100925.1$+$325553 at
$z=0.508$, though no member galaxies have spectroscopic redshifts,
most of the luminous member galaxies ($M_r\le-21$) of these clusters
can be well discriminated by using photometric redshifts. For cluster
WHL J145044.4$+$220134 at $z=0.601$, 18 luminous red galaxies are
discriminated. Some probable cluster galaxies are not selected as
members because of poor estimate of photometric redshift at $z\sim0.6$
(see Figure~\ref{phzg}). In general, these examples show that
photometric redshifts can be very efficient indicator for picking up
cluster galaxies up to $z\sim0.5$ in the SDSS, much deeper than that
by spectroscopic redshifts.

%%%%%%%%%%%%%%%%%%%%%%%%%%%%%%%%%%%%%%%%%%%%%%%%%%%%%%%%%%%%%%%%%%%%%%%%5
\section{Statistical tests for the identified Clusters}

Using the SDSS spectroscopic redshifts, we estimate the uncertainty of
cluster redshift, the contamination rate, and the completeness of
discriminated member galaxies. We also examine the reliability of
cluster richness determined by our method.
Moreover, Monte Carlo simulations are performed with the real observed
background of galaxies to estimate cluster detection rate and false
detection rate of our algorithm.

\subsection{Redshift test}

\begin{figure}
\epsscale{1.}
\plotone{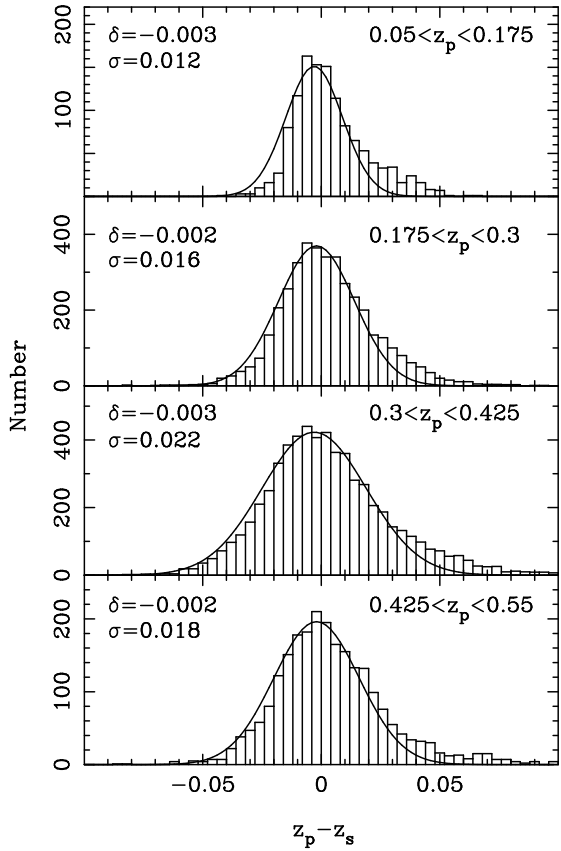}
\caption{Distribution of the difference between photometric and
  spectroscopic redshifts in the four redshift ranges. The solid line
  is the best fit with a Gaussian function. The parameters, i.e., the
  offset $\delta$ and the standard deviation $\sigma$, of the Gaussian
  function are marked on the left of each panel.
\label{phzc}}
\end{figure}

We verify the accuracy of photometric redshifts of clusters in
our catalog. The spectroscopic redshift of a cluster is taken to be
that of its brightest cluster galaxy (BCG). From the SDSS data, we
find BCGs of 13,643 clusters having spectroscopic redshifts measured.
In Figure~\ref{phzc}, we show the distribution of difference between
photometric and spectroscopic redshifts, $z_p-z_s$. In each panel,
we fit the distribution of $z_p-z_s$ with a Gaussian function. The
systematic offset $\delta$ of the fitting is $-0.002$ or $-0.003$, and
the standard deviation $\sigma$ is around 0.02.

\subsection{Member detection and richness tests}
\label{concomp}

\begin{figure}
\epsscale{1.}
\plotone{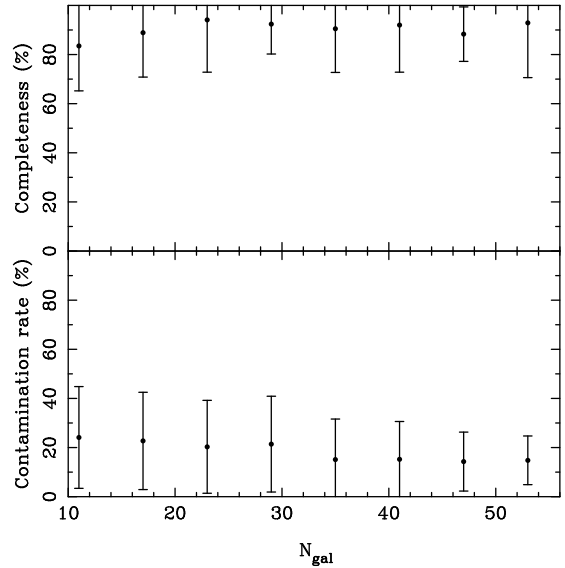}
\caption{Completeness (the upper panel) and contamination rate (the 
    lower panel) of member galaxy candidates within a radius of 1 Mpc
  against the number of member galaxy candidates, $N_{\rm gal}$.
\label{memb}}
\end{figure}

Within photometric redshift gap, member galaxies can be contaminated
by foreground and background galaxies and incompletely detected. 
Now, we use the spectroscopic redshifts of the SDSS DR6 to study the
contamination due to projection effect and the completeness of member
galaxies discrimination. From our sample, we obtain 1070 clusters with
more than five discriminated members having spectroscopic
redshifts. Totally, 10,677 galaxies with spectroscopic redshifts are
discriminated as members of these clusters. The cluster redshift can
be defined to be the median redshift of these member galaxies with
spectroscopic redshifts for each of these clusters. We compare
individual member redshifts with the estimate of the cluster redshift,
and find that 2260 (21\%) galaxies have velocity difference from
clusters by more than 2000 km s$^{-1}$. They are probably not the
member galaxies, and therefore are considered as contamination of
galaxies.
However. this percentage is somewhat biased by the SDSS spectroscopic
selection.  Since bright galaxies are preferentially targeted in the
SDSS spectroscopic survey, the spectroscopically measured galaxies in
the dense region are more likely member galaxies of clusters. Assuming
that the effect of other selection bias, e.g., fiber collisions, is
limited on measurement of galaxies, the fraction of 21\% can be
considered as a lower limit of contamination rate.

To estimate the completeness, we get the total member galaxies in the
SDSS data. In the 1070 clusters, 8793 galaxies of $M_r\le-21$ within 
1 Mpc from cluster centers have velocities differing from
those of clusters by less than 2000 km s$^{-1}$, of which 7882 (90\%)
galaxies have been found as member galaxies by our method.
Figure~\ref{memb} shows the contamination rate and the completeness of
member galaxy candidates against the number of member galaxy
candidates. The completeness of member galaxies is nearly constant
for clusters with different richnesses. The contamination rate is
roughly 20\%, and slightly decreases with $N_{\rm gal}$.

\begin{figure}
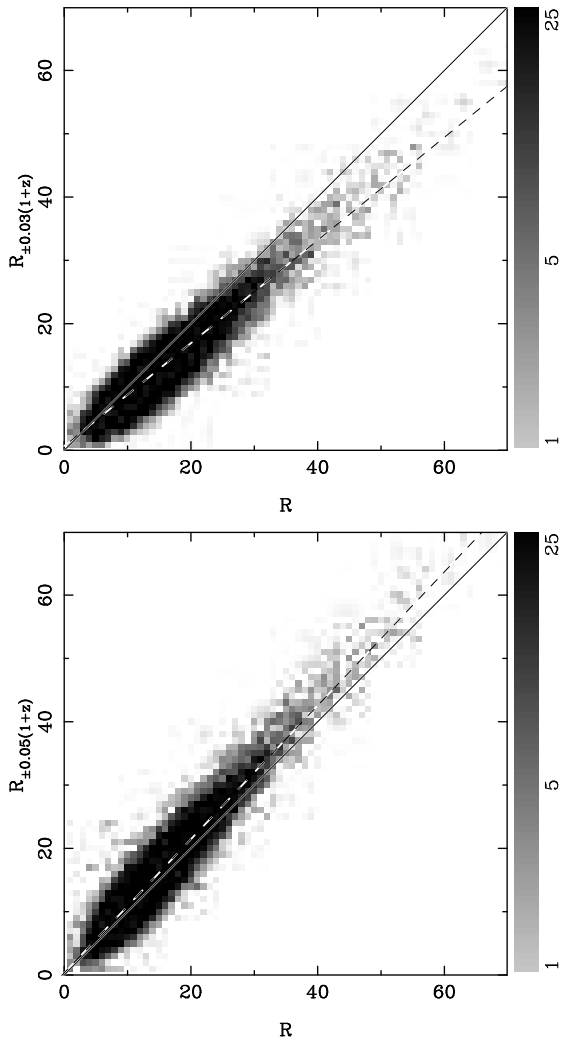

\epsscale{1.}
\plotone{f9a.eps}\\[2mm]
\plotone{f9b.eps}
\caption{Comparison between the cluster richness for the gap
  of $z\pm0.04(1+z)$ and that for the gap of $z\pm0.03(1+z)$ (the 
    upper panel) and $z\pm0.05(1+z)$ (the lower panel). The dashed
  line shows the best linear fit. The solid line is an equal line.
\label{richtest}}
\end{figure}

The contamination rate and the completeness depend on photometric redshift
gap. With a larger gap, real member galaxies are selected more
completely, but the member contamination becomes severer.  With a
smaller gap, we can discriminate less real member galaxies with a
small contamination, but the sensitivity of cluster detection is
lower. The gap about $z\pm0.04(1+z)$ is a reliable compromise, within
which the majority of member galaxies in a cluster can be picked out
with only a small percentage of contamination galaxies included (see
Figure~\ref{memb}).  To study how the richness depends on the gap, we
obtain richnesses of clusters using different photometric redshift
gaps. Figure~\ref{richtest} compares these cluster richnesses. They
are tightly correlated with relations of
\begin{equation}
R_{\pm0.03(1+z)}=(0.67\pm0.01)+(0.81\pm0.01)\times R,
\end{equation}
and
\begin{equation}
R_{\pm0.05(1+z)}=(0.33\pm0.01)+(1.06\pm0.01)\times R.
\end{equation}
Statistically, the tight correlations suggest that any richness within
a gap between $z\pm0.03(1+z)$ and $z\pm0.05(1+z)$ can be an equivalent
indicator of true richness. The richness does not change much for that
with the gap of $z\pm0.05(1+z)$, indicating that member galaxies are
selected with a good completeness for the gap of $z\pm0.04(1+z)$.

\subsection{Cluster detection rate}
\label{comprate}

Mock clusters are simulated with assumptions for their distributions 
and then added to the real data of the SDSS to test the detection rate
of the mock clusters by our cluster-finding algorithm.

The luminosity function of galaxies in a cluster is taken to follow
the Schechter function \citep{sch76}
\begin{equation}
\phi(M)dM\propto 10^{-0.4(M-M^{\ast})(\alpha+1)}\exp[-10^{-0.4(M-M^{\ast})}]dM.
\end{equation}
We adopt the parameters as, $\alpha=-0.85\pm0.03$,
$M^{\ast}=-22.21\pm0.05$, derived by \citet{gom+02} based on the SDSS
CE clusters. We also assume that the galaxy number density in a mock
cluster follows the Navarro-Frenk-White (NFW) profile \citep{nfw97}, in which the scaled
radius is adopted to be 0.25 Mpc \citep[for clusters with masses of
  $\sim$$10^{14}$~$M_{\odot}$; see][]{pap05}.
The mock clusters are distributed in redshift space with a uniform
comoving number density.  Then, we calculate the apparent magnitudes
of cluster galaxies after correcting their colors in the $r$ band with
the $K$-correction curve of early-type galaxy by \citet{fsi95}. The
photometric redshifts are assigned to the member galaxies of each
cluster. We assume that the uncertainty of photometric redshift of
cluster galaxies follows a Gaussian probability function with a
standard deviation of $\sigma_z$, but varies with redshift in the form
of $\sigma_z=\sigma_0(1+z)$.

\begin{figure}
\epsscale{1.}
\plotone{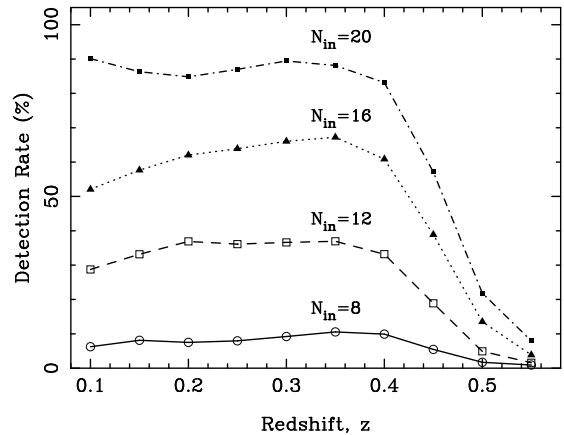}\\[2mm]
\caption{Detection rate of mock clusters as a function of redshift in
  the real background for input richness of $N_{\rm in}=8$ (the open
  circle), 12 (the open square), 16 (the black triangle), and 20 (the black
  square) assuming $\sigma_z=0.03(1+z)$.
\label{recov1}}
\end{figure}

We do two tests. First, we test with mock clusters for different given
richness independently. Here, the input richness for a mock cluster,
$N_{\rm in}$, is defined to be the numbers of luminous galaxies
($M_r\le-21$) within a radius of 1 Mpc. For each richness, 2000 clusters
are simulated in the redshift range $0.05<z<0.55$ via above procedures
assuming $\sigma_z=0.03(1+z)$, and added to the real SDSS data of 500
deg$^2$. Mock clusters are put to a region where no real detected
cluster exists within 3 Mpc. Our cluster finding algorithm is then
performed to detect these mock clusters from the galaxy sample of
$r\le21.5$.

A mock cluster is detected if the number of recognized member galaxies
($M_r\le-21$) is above the detection threshold of eight within a radius of
0.5 Mpc and the redshift gap (see Section~\ref{algorithm}). Here, the
recognized members can be not only the member galaxies of mock
clusters, but also the contamination galaxies from the real
background. We emphasize again that the input richness $N_{\rm in}$
and output richness $R$ is for a radius of 1 Mpc, but the detection
threshold is designed within a radius of 0.5 Mpc. Figure~\ref{recov1}
shows the detection rates as a function of redshift for mock clusters
with different input richness.  The detection rates depend on input
richness, but do not vary much with redshift at $z<0.4$. The detection
rates of clusters with input richness of $N_{\rm in}=8$ (the open circles)
are about 10\% up to $z\sim 0.4$. The detection rates increase to 35\%
for clusters of $N_{\rm in}=12$ (the open square), and more than 60\% for
clusters of $N_{\rm in}=16$ (the black triangle) and 90\% for $N_{\rm
  in}=20$ (the black square) up to $z\sim 0.4$. The detection rates
decrease at a higher redshift due to the magnitude cutoff, as
mentioned in Section~\ref{algorithm}.

\begin{figure}
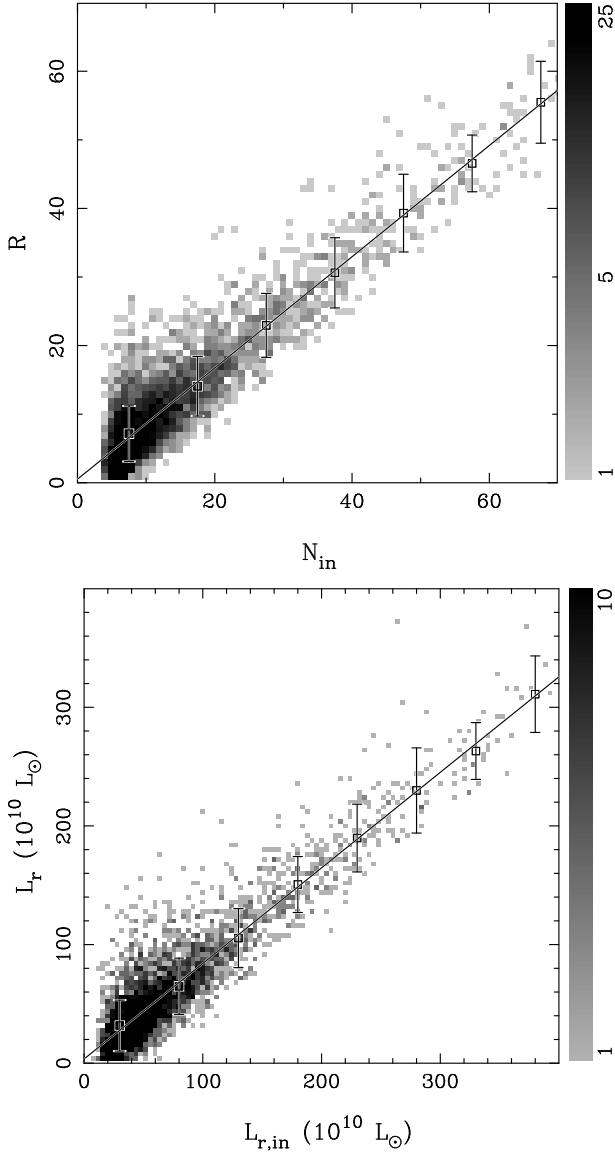

\epsscale{1.1}
\plotone{f11a.eps}\\[2mm]
\plotone{f11b.eps}
\caption{Comparison between the input and output richness (the upper
    panel), and the input and output summed luminosity (the lower
    panel) for clusters of $z<0.42$.
\label{simcom3}}
\end{figure}

\begin{figure}
\epsscale{1.}
\plotone{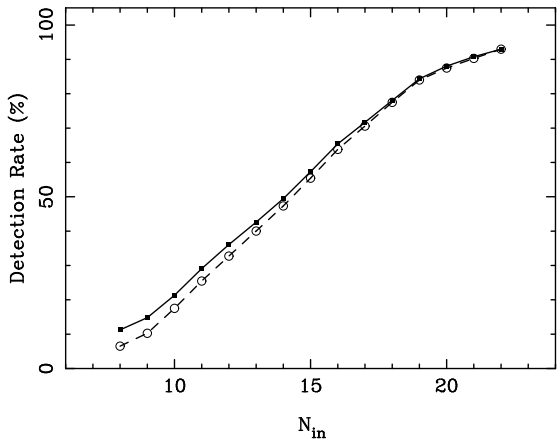}
\caption{Detection rate as a function of input richness at $z<0.42$
  for all detected clusters (the solid line) and more restrictively
  excluding those with more than half contamination (the dashed line).
\label{recov2}}
\end{figure}

\begin{figure}
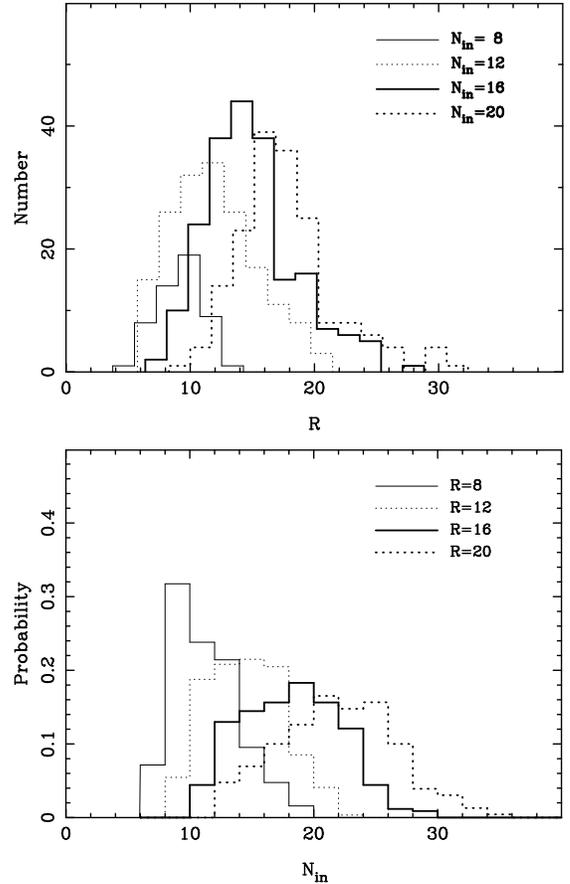

\epsscale{1.}
\plotone{f13a.eps}\\[2mm]
\plotone{f13b.eps}
\caption{Upper panel: Output richness distribution of detected
  clusters for different input richness. Lower panel:
  Probability distribution of input richness for clusters that have
  the same output richness.
\label{histout}}
\end{figure}

\begin{figure}
\epsscale{1.}
\plotone{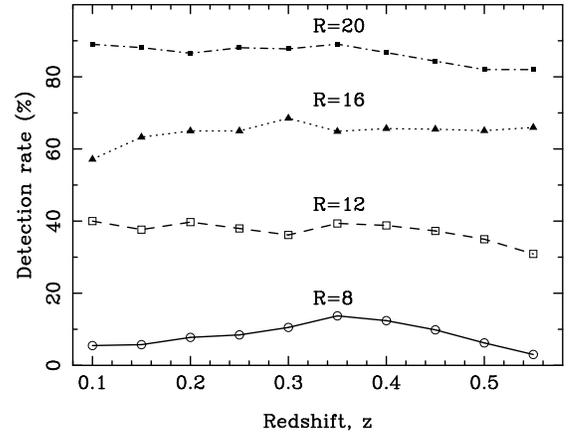}
\caption{Detection rate of mock clusters as a function of redshift
  for different output richnesses.
\label{simcopl3}}
\end{figure}

Secondly, we perform Monte Carlo simulation considering a population
of clusters with various input richness. Using the mass function of
\citet{jfw+01} in a cosmology with $\Omega_m=0.3$ and $\sigma_8=0.9$,
we generate the halos with masses greater than $10^{14}~M_{\odot}$ in
the redshift range $0.05<z<0.55$.
According the halo occupation distribution obtained by \citet{ymv+08},
we derive the number of the galaxies of $^{0.1}M_r-5\log h\le-20$ in
the halos.  Here, $^{0.1}M_r$ refers to the absolute magnitude
$K$-corrected and evolution corrected to $z=0.1$ in the $r$-band.  The
magnitudes, coordinates, redshifts and cluster input richness are
simulated as described above. These mock galaxies are added to the
real background, and then we detect them using our cluster-finding
algorithm.

For every input cluster, we discriminate the luminous ``member
galaxies'' ($M_r\le-21$) by using photometric redshifts and obtain
the output richness. We also estimate its luminosity by {\it summing
  luminosities of ``member galaxies'' after contamination
  subtraction}. In Figure~\ref{simcom3}, we compare the input and
output richnesses and the summed luminosities for clusters of
$z<0.42$.  The output richness is well related to the input richness
with a scatter of $\sim$5. The best fit gives
\begin{equation}
R=(0.52\pm0.81)+(0.81\pm0.01)N_{\rm in}.
\label{rinout}
\end{equation}
Similar, the output summed luminosity is also well related to the
input luminosity a scatter of $\sim$30$\times10^{10}~L_{\odot}$. The
best fit gives
\begin{equation}
L_{r,10}=(3.58\pm2.39)+(0.81\pm0.01)L_{r,\rm in, 10}.
\end{equation}
Here $L_{r,10}$ refers to the summed $r$-band luminosity in unit of
$10^{10}~L_{\odot}$.

Again, a mock cluster is detected by our algorithm if more than eight
luminous ``member galaxies'' are found within a radius of 0.5
Mpc. Figure~\ref{recov2} shows the detection rate as a function of
input richness for clusters of $z<0.42$. The detection rate is 10\%
for clusters of $N_{\rm in}=8$ if all detected clusters are
considered. However, if the number of member candidates of a detected
cluster is more than twice of the input richness, then more than half
member candidates are contamination galaxies. One can consider it as a
false detection of a cluster. The detection rate becomes 6\% if more
restrict criterion for cluster detection is applied. The detection rate
reaches 60\% for clusters of $N_{\rm in}=16$, and 90\% for clusters of
$N_{\rm in}=20$, respectively.

In Figure~\ref{histout}, we show the output richness distribution of
detected clusters for different input richness and the probability
distribution of input richness for clusters that have the same output
richness. Clusters with larger output richnesses are from larger input
clusters. About 80\% of detected clusters of $R=12$ are mock clusters
of $N_{\rm in}\le16$, while about 70\% detected clusters of $R=16$ are
mock clusters of $N_{\rm in}\ge16$.
The detection rates by our algorithm for different output richnesses
are shown in Figure~\ref{simcopl3}. The detection rates are $\sim$40\%
for clusters of $R=12$, which increase to $\sim$60\% for clusters
of $R=16$ and $\sim$90\% for clusters of $R=20$.
As one can see from Figure~\ref{simcom3}, clusters with input richness
$N_{\rm in}\le12$ can have output richness $R\ge16$ due to
contamination from real background. Since there are significantly more
relatively poorer clusters than big ones, many clusters of $R\ge16$ in
the output catalog would be poor ones if the detection threshold
(i.e., eight galaxies within a radius of 0.5 Mpc) is not used. Our
algorithm preferentially detects the rich clusters as shown above,
and hence reduces the contamination from the poor clusters in the output
catalog. The above simulations show that the completeness of cluster
detection by our method is nearly constant up to $z\sim0.42$ using
photometric redshift catalog of the SDSS. The output catalog is
$\sim$60\% complete for clusters with $N_{\rm in}=16$, and $\sim$90\%
complete for clusters with $N_{\rm in}=20$.

\subsection{False detection rate}
\label{falsrate}

The presence of the large-scale structures makes it possible to detect
false clusters because of projection effect. We also perform Monte
Carlo simulation with the real SDSS data to estimate the false
detection rate. Our method is similar to that of 
\citet{gsn+02}. First, each galaxy in the real SDSS data is forced to
have a random walk in the two-dimension projected space in a random
direction. The step length is a random value less than 2.5
Mpc. Second, we shuffle the photometric redshift of the galaxy
sample. The procedures above are to eliminate the real clusters, but
reserve the larger scale structure in two-dimension projected
space. The maximum step of 2.5 Mpc is chosen so that clusters as rich
as $N_{\rm gal}=100$ can be eliminated. Our method is applied to
detect ``clusters'' from the shuffled sample of 500 deg$^2$.

\begin{figure}
\epsscale{1.}
\plotone{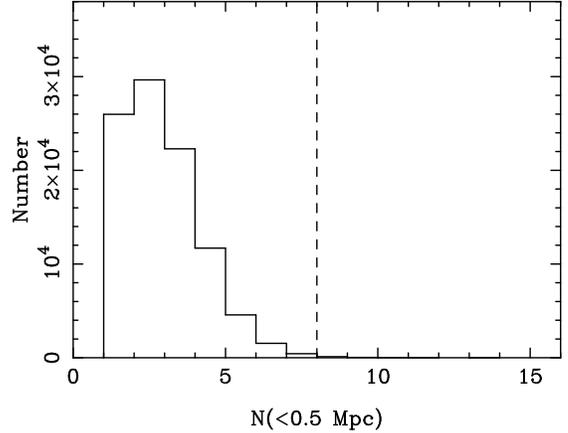}
\caption{Distribution of the number of galaxies ($M_r\le-21$) within
  0.5 Mpc from ``cluster center'' and a redshift gap between
  $z\pm0.04(1+z)$ in the shuffled data. The dashed line represents the
  threshold to identify clusters.
\label{false}}
\end{figure}

\begin{figure}
\epsscale{1.}
\plotone{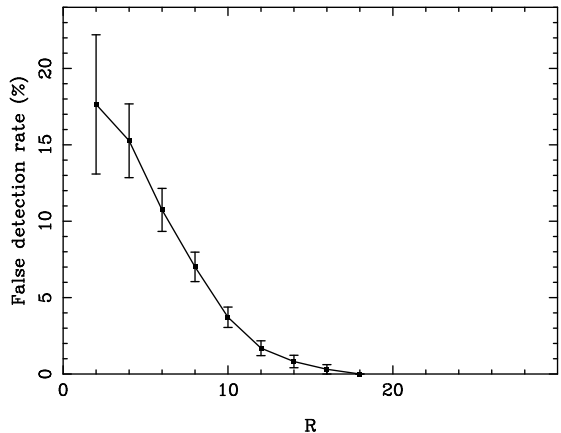}
\caption{False detection rate as a function of cluster richness.
\label{falngal}}
\end{figure}

Figure~\ref{false} shows the distribution of the number counts of
galaxies ($M_r\le-21$) within 0.5 Mpc from the centers of
``cluster'' candidates and the photometric redshift gap of
$z\pm0.04(1+z)$.  Only 148 ``clusters'' are found to exceed the
threshold (the dashed line), comparing to the 2380 real detected clusters
in the 500 deg$^2$ region. We cross-identify the ``clusters'' with
real clusters within a radius of 1 Mpc, and find that 41 ``clusters''
match the real clusters, which means that they are clusters not well
shuffled to a good randomness. The rest 107 clusters are considered as
false clusters. This simulation shows that our algorithm gives a false
detection rate of clusters as $107/2380\simeq$5\%. The rate decreases
with increase in the cluster richness as shown in Figure~\ref{falngal}.
We also take the maximum step length of 4 Mpc, and the false detection
rate becomes $72/2380\simeq$3\%.

%%%%%%%%%%%%%%%%%%%%%%%%%%%%%%%%%%%%%%%%%%%%%%%%%%%%%%%%%%%%%%%%%%
\section{Comparison with previous optical-selected cluster catalogs}

We compare our cluster catalog with previous catalogs, the Abell, the
SDSS CE, and the maxBCG catalogs. The Abell catalog contains most of
rich clusters at $z<0.2$ but without a quantitative measurement of
completeness \citep{aco89}. The SDSS CE catalog contains poor clusters
as well as rich ones at $z<0.44$ \citep{gsn+02}. The SDSS maxBCG catalog
has a uniform selection in the redshift range $0.1<z<0.3$
\citep{kma+07b}.

\subsection{Comparison with the Abell clusters}

\begin{figure}
\epsscale{1.}
\plotone{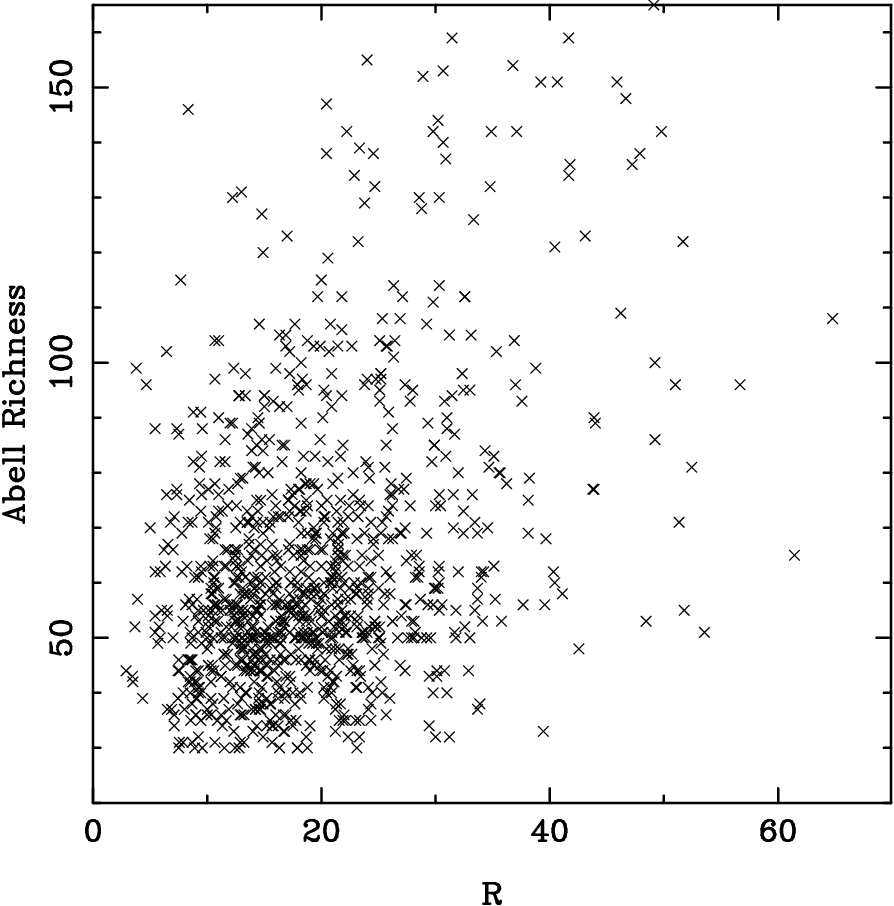}\\[2mm]
\plotone{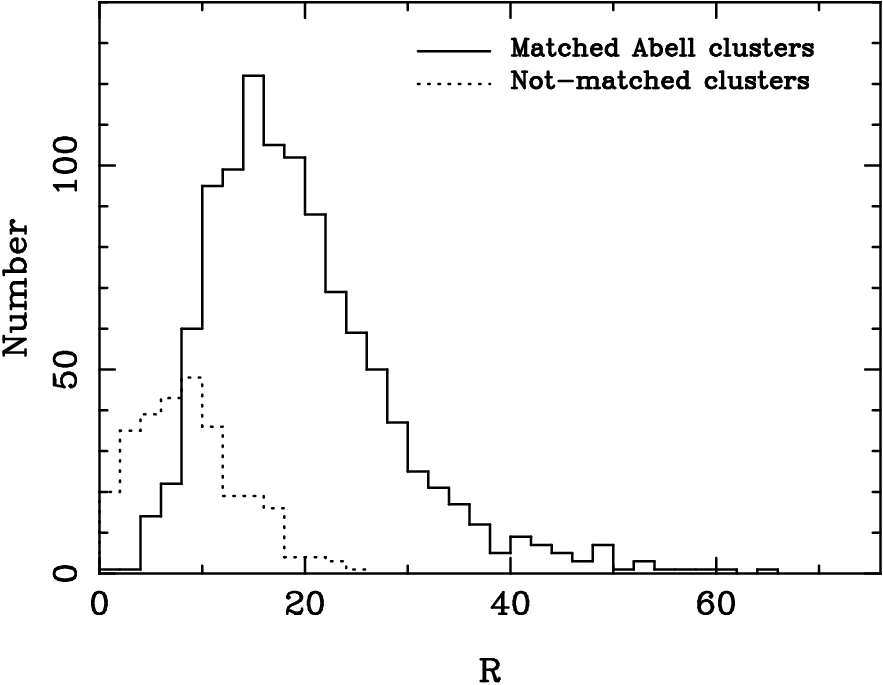}
\caption{Upper panel: Comparison between the Abell richness and the
  richness we determine for the matched clusters. Lower panel:
  Distributions of richness for the matched and not-matched clusters.
\label{rich_abell}}
\end{figure}

There are 1594 Abell clusters in the sky region of the SDSS DR6. Some
Abell clusters have redshifts not measured previously. We take their
redshifts to be the values of the BCGs from the SDSS data. The
photometric redshifts are used if no spectroscopic redshifts are
available. Totally, 1354 Abell clusters have redshifts $z>0.05$, of
which 991 clusters are found within a projected separation of $r_p<1$
Mpc and redshift difference of $\Delta z<0.05$ (about 2.5\,$\sigma$ of
our cluster redshift accuracy) from clusters in our catalog. Another
53 Abell clusters are found within a projected separation of
$1.0<r_p<1.5$ Mpc and redshift difference of $\Delta z<0.05$ from
clusters in our catalog, which are likely to have substructures so that
centers are defined at different substructures in two catalogs. In
total, 1044 (77\%) Abell clusters are considered to be matched with
our catalog.

In Figure~\ref{rich_abell}, we compare the Abell richness with the
richness we determine for the matched clusters. The correlation is
poor. The discrepancy may come from the uncertainties of the
Abell richness. \citet{yl99} showed that the Abell richness for
clusters of $z\ge0.1$ is not a good indicator of their true richness,
and sometimes the richness is overestimated by as much as a factor of
3. One reason is the Abell richness suffers from the projection
effect. Simulation shows that cluster surveys in two dimensions are
heavily contaminated by projection biases if the cluster search radius
is as large as Abell radius of 1.5 $h^{-1}$~Mpc \citep{vfw97}.
Another reason for the null correlation may be the uncertainty of the
definition. Recall that the Abell richness is defined to be the number
of galaxies within 2 mag range below the third-brightest galaxy within
Abell radius after correcting background. The richnesses are
calculated within various absolute magnitude range because the
magnitudes of the third-brightest galaxies vary a lot.
For the non-matched Abell clusters, we also determine their richnesses
by our method. The matched Abell clusters have a high richness, while
the not-matched clusters are relatively poor with richness around eight,
few larger than 20 (see the lower panel of Figure~\ref{rich_abell}).

\subsection{Comparison with the SDSS CE clusters}

\begin{figure}
\epsscale{1.}
\plotone{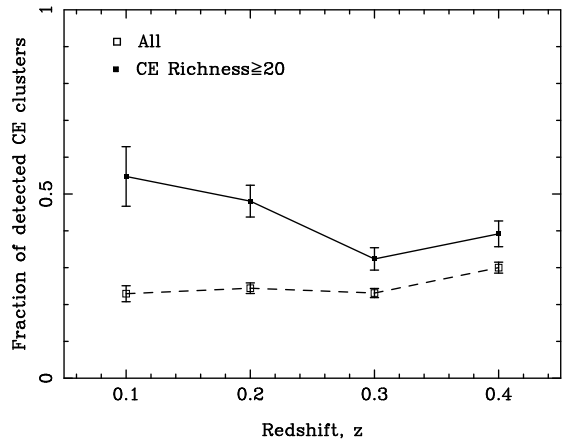}
\caption{Fraction of the CE clusters we detected as a function of
  redshift.
\label{rate_ce}}
\end{figure}

\begin{figure}
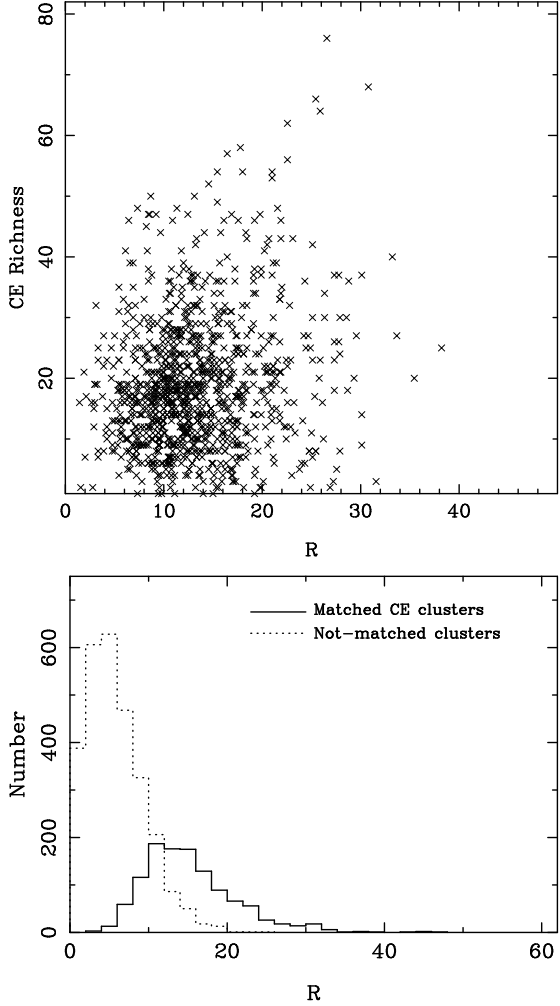

\epsscale{1.}
\plotone{f19a.eps}\\[2mm]
\plotone{f19b.eps}
\caption{Same as Figure~\ref{rich_abell} but for the CE clusters.
\label{rich_ce}}
\end{figure}

The SDSS CE clusters were identified by using 34 color cuts. The
redshifts of clusters were estimated with the uncertainties of
$\sigma=0.0147$ at $z<0.3$ and $\sigma=0.0209$ at $z>0.3$. The
CE richness is defined to be the number of galaxies within 2 mag range
below the third-brightest galaxy and within the detection radius after
correcting background \citep{gsn+02}.

Among 4638 CE clusters, 1160 clusters are found within a projected
separation of $r_p<1.5$ Mpc and redshift difference of $\Delta z<0.05$
from clusters in our catalog. Figure~\ref{rate_ce} shows the detection
rates of the CE clusters by our method as function of redshift. The
rates are about 20\%--30\% for the whole sample, and increase to
40\%--50\% for clusters with the CE richness $\ge20$.
The correlation between our richness and the CE richness is also poor 
(see Figure~\ref{rich_ce}), suggesting that the CE richness has a
large uncertainty. For the not-matched CE clusters, we determine their
richness by our method and find that most of the not-matched
clusters are relatively poor with mean richness $\sim$3 (see the lower
panel of Figure~\ref{rich_ce}). Obviously, the CE clusters we detected
are much richer than the not-matched clusters.

\subsection{Comparison with the SDSS maxBCG clusters}

\begin{figure}
\epsscale{1.}
\plotone{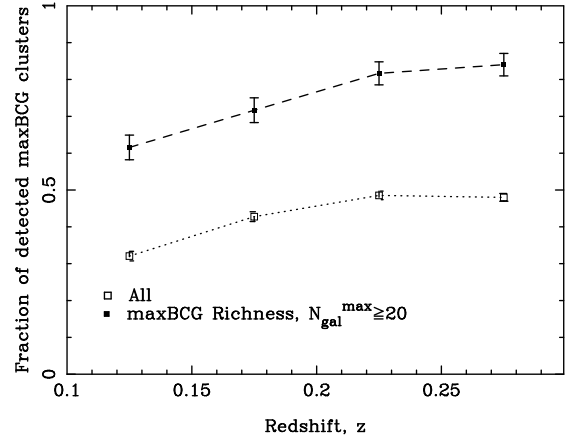}
\caption{Same as Figure~\ref{rate_ce} but for the maxBCG clusters.
\label{rate_max}}
\end{figure}

\begin{figure}
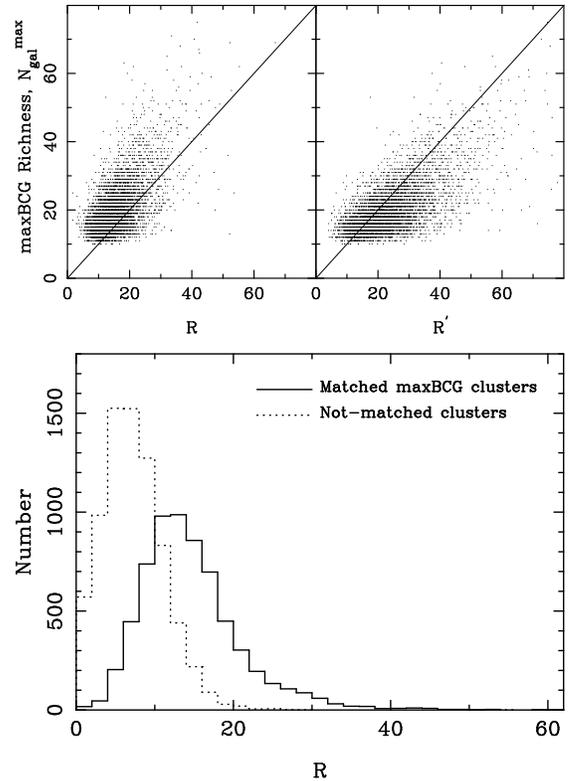

\epsscale{1.}
\plotone{f21a.eps}\\[2mm]
\plotone{f21b.eps}
\caption{Same as Figure~\ref{rich_abell} but for the maxBCG clusters.
  $R'$ refers to cluster richness by our method using the
  same radius and magnitude range with the maxBCG clusters. The solid
  line is an equal line.
\label{rich_max}}
\end{figure}

The SDSS maxBCG is approximately 85\% complete in the redshift range
$0.1<z<0.3$ with masses  $M>1\times10^{14}~M_{\odot}$ \citep{kma+07b}. The
redshifts of clusters were estimated with the uncertainties of
$\sigma=0.01$. The cluster richness, $N_{\rm gal}^{\rm max}$, is
defined to be the number of galaxies within a radius of 1 $h^{-1}$ Mpc
and $2\,\sigma_c$ of the ridgeline colors, brighter than
$0.4L^{\ast}$.
A scaled richness, $N_{200}$, is measured to be the number of
galaxies within $r_{200}$ and the color cuts. Here, $r_{200}$ is the
radius within which the mean mass density is 200 times that of the
critical cosmic mass density.
Among 13,823 maxBCG clusters, 6424 clusters are found within a
projected separation of $r_p<1.5$ Mpc and redshift difference $\Delta
z<0.05$ from the clusters in our catalog. As shown in
Figure~\ref{rate_max}, the detection rates of the maxBCG clusters by
our method are 40\%--50\% for the whole sample, and increase
70\%--80\% for clusters with the maxBCG richness $\ge20$.

The luminosity cutoff, $0.4L^{\ast}$, of the maxBCG method corresponds
to absolute magnitude of $M_r\simeq-20.6$ \citep{kma+07b}, which is
about 0.4 magnitude fainter than that of our work. To make a
comparison, we calculate the richness, $R'$, for the matched clusters
within the same radius and magnitude range with the maxBCG clusters,
i.e., 1 $h^{-1}$~Mpc and $M_r\le-20.6$.
In Figure~\ref{rich_max}, we compare the maxBCG richness, $N_{\rm
  gal}^{\rm max}$, with the richnesses, $R$ and $R'$. Both
correlations are tighter than those with the Abell and CE clusters,
though large scatters exist.
With the same selection criteria, we find that the maxBCG richness is
systematically smaller than the richness by our method. The discrepancy
may come from some systematic bias in maxBCG method. Recall that the
maxBCG method only selects ridgeline member galaxies without
contamination subtraction. The color--magnitude diagrams
\citep[e.g.,][]{mnr+05} show that many member galaxies fall outside of
the ridgeline of red galaxies, and hence they are likely missed by the
maxBCG method. \citet{rrk+08} also pointed out the systematic bias
for the maxBCG richness due to color off-sets. The ridgeline galaxies
fall outside the color cuts because of the increasing 
photometric errors with redshift. In addition, the ridgeline of red
galaxies is not as flat as assumed and even evolves with redshift, so
that the color cuts based on the BCGs colors will lose some of the
less luminous cluster member galaxies.
Furthermore, the study by \citet{dsm+02} shows that some massive
clusters do not have a prominent red sequence, which could induce bias
in cluster detection and richness measurement.

We show that our method tends to detect the rich maxBCG clusters and
that the not-matched maxBCG clusters are relatively poor with mean
richness $\sim$6 (see the lower panel of Figure~\ref{rich_max}).

%%%%%%%%%%%%%%%%%%%%%%%%%%%%%%%%%%%%%%%%%%%%%%%%%%%%%%%%%%%%%%%%%%
\section{Correlations of our clusters with X-ray Measurements}
\label{cluster_xray}

The measurements in X-rays provide the properties of clusters from hot
intracluster gas. The imaging observations can give the X-ray
luminosity, and spectroscopic observations can provide the temperature
of hot gas. Using the measurements in X-rays, the gravitational
cluster mass can be derived \citep{wu94,rb02}.
With luminous member galaxies well discriminated, the correlations
between these X-ray measurements and the cluster richness or the summed
optical luminosities are expected.

\begin{figure}
\epsscale{1.}  
\plotone{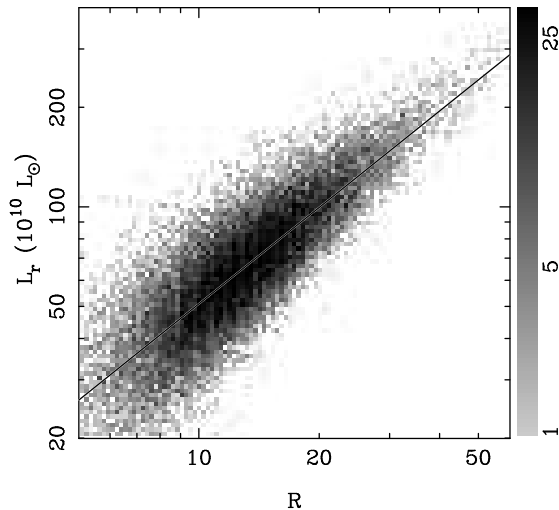}
\caption{Correlation between the richness and the summed luminosity
  for clusters of $z\le0.42$. The solid line is the best power-law
  fit given in Equation~(\ref{rl}).
\label{rich_lum}}
\end{figure}

As mentioned above, the faint end of member galaxies is
$M_r=-21$ in our sample at $z\le0.42$. At higher redshifts, the faint
end moves to a brighter magnitude depending on redshift, so that the
estimated summed luminosities for clusters of $z>0.42$ are
biased. Therefore, we only consider clusters of $z\le0.42$ in the
following statistics.

Figure~\ref{rich_lum} shows the correlation between the richness and
the summed luminosities for clusters of $z\le0.42$.  We find that the
summed luminosity of a cluster is linearly related to the cluster
richness by
\begin{equation}
L_{r,10}=(5.47\pm0.06)R^{0.97\pm0.01}.
\label{rl}
\end{equation}
This is consistent with the relation found by \citet{pbb+07}.

\begin{figure}
\epsscale{1.1}
\plotone{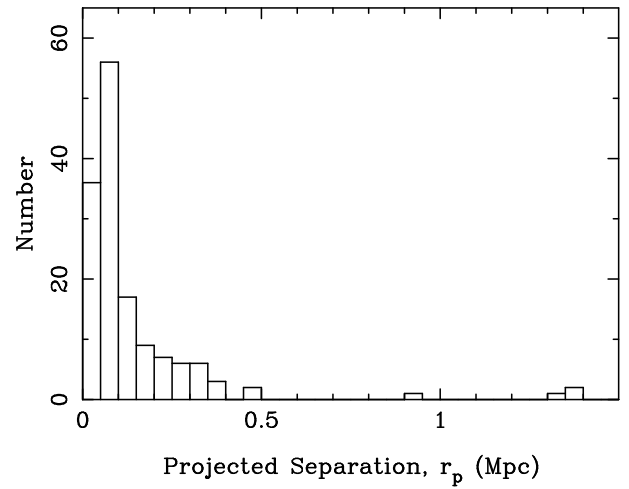}
\caption{Distribution of projected separations between X-ray peaks of
  clusters and the optical BCGs.
\label{lxd}}
\end{figure}

\begin{figure}
\epsscale{1.1}
\plotone{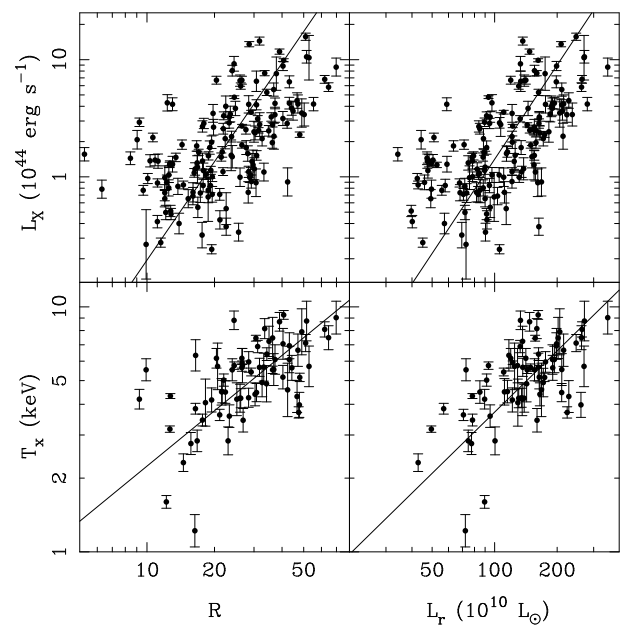}
\caption{Correlations between the cluster richness ($R$,
  the left panels) and summed $r$-band luminosity ($L_r$, 
    the right panels) with the X-ray luminosity (the upper
    panels) of 146 clusters from \citet{bvh+00,bsg+04}, and
  temperature (the bottom panels) of 67 clusters from
  \citet{fmo04}. The lines are the best fit given in Equations~(\ref{lxr})--(\ref{txl}).
\label{lum_lx}}
\end{figure}

\subsection{Correlations between the richness and optical luminosity with the X-ray luminosity and temperature}
\label{opt_xray}

There are 239 (203 NORAS and 36 REFLEX) X-ray clusters from
\citet{bvh+00,bsg+04} in the sky region of the SDSS DR6, of which 190
clusters have redshifts $z>0.05$. We find 146 {\it ROSAT} X-ray clusters
within a projected separation of $r_p<1.5$ Mpc and the redshift
difference of $\Delta z<0.05$ from clusters in our sample. The X-ray
emission of clusters usually traces the centers of matter
distributions. They are likely coincident with the BCGs probably
located near the centers of clusters. Figure~\ref{lxd} shows the
distribution of the projected separation between X-ray peaks of clusters
and the optical BCGs. Most (132/146) of the clusters have a separation
of $r_p\le0.3$ Mpc.
The small offsets are probably due to the movement of BCGs with respect to
the cluster potential \citep{oh01}.
Five merging clusters have projected separations of $r_p\ge0.5$ Mpc
because the BCGs and X-ray peak are located at different subclusters.

We find that the richness and summed $r$-band luminosities of 146
X-ray clusters are well correlated with the X-ray luminosity
\citep{bvh+00,bsg+04} derived from the {\it ROSAT} observations (see
Figure~\ref{lum_lx}). The best fit to the data gives
\begin{equation}
\log L_{X,44}=(-3.50\pm0.17)+(2.79\pm0.13)\log R,
\label{lxr}
\end{equation}
and
\begin{equation}
\log L_{X,44}=(-5.19\pm0.25)+(2.67\pm0.12)\log L_{r,10}.
\end{equation}
where $L_{X,44}$ refers to X-ray luminosity in the 0.1--2.4 keV band
in unit of $10^{44}~{\rm erg~s^{-1}}$. The tight correlations suggest
that the member galaxies are well discriminated by our method, as
shown in Section~\ref{concomp}.
\citet{pbb+05} studied the correlations between the optical and X-ray
measurements using the RASS--SDSS clusters. They obtained the slope of
the $L_{X,44}$--$L_{r,10}$ relation to be $1.72\pm0.09$, much smaller than 
our result.
Using maxBCG clusters, \citet{rmb+08} studied the mean and scatter of
the $L_{X,44}$--$N_{200}$ relation, and obtained the slope of
$1.82\pm0.05$, where $N_{200}$ are determined within different radius,
$r_{200}$, for different clusters. To make a comparison, we scale the
slope of their relation to that of $L_{X,44}$--$N_{\rm gal}^{\rm
  max}$, where $N_{\rm gal}^{\rm max}$ is also defined within a fixed
radius. The scaling relation between $N_{200}$ and $N_{\rm gal}^{\rm
  max}$ is $N_{200}\propto (N_{\rm gal}^{\rm max})^{1.41\pm0.01}$
\citep{kma+07a}. Therefore, the $L_{X,44}$--$N_{\rm gal}^{\rm max}$
relation has a slope of 2.57$\pm$0.09, in agreement with our result.
However, the correlations of the $L_{X,44}$--$R$ relation are much
tighter than that shown using the maxBCG clusters \citep[see Figure 7
  of][]{rmb+08}.

\citet{fmo04} compiled the temperatures of $\sim$300 X-ray
clusters, of which 67 clusters are found in our catalog. We plot
the richness and the summed $r$-band luminosity against the X-ray
temperature for 67 clusters in Figure~\ref{lum_lx} and find the
best fit as
\begin{equation}
\log T_X=(-0.40\pm0.12)+(0.75\pm0.08)\log R,
\end{equation}
and
\begin{equation}
\log T_X=(-1.09\pm0.18)+(0.83\pm0.08)\log L_{r,10},
\label{txl}
\end{equation}
where $T_X$ refers to X-ray temperature in unit of keV. The slope of
the $T_X$--$L_r$ relation is slightly higher than $0.61\pm0.03$ found
by \citet{pbb+05}.

\begin{figure}
\epsscale{1.1}
\plotone{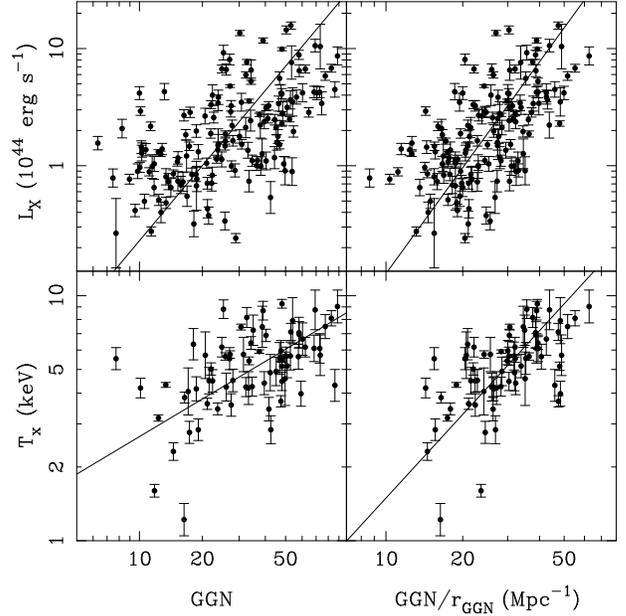}
\caption{Correlations between the cluster $GGN$ (the left panels)
  and $GGN/r_{\rm GGN}$ (the right panels) with the X-ray
  luminosity (the upper panels) and temperature (the
    bottom panels). The lines are the best fit.
\label{ggn_lx}}
\end{figure}

We also find the correlations between the $GGN$ and $GGN/r_{\rm GGN}$
of clusters with the X-ray luminosity and temperature.
Figure~\ref{ggn_lx} shows the correlations. $GGN/r_{\rm GGN}$
is more tightly correlated with $L_{X,44}$ and $T_X$ than $GGN$. The 
$L_{X,44}$--$GGN/r_{\rm GGN}$ relation and the $T_X$--$GGN/r_{\rm
  GGN}$ relation are:
\begin{equation}
\log L_{X,44}=(-3.92\pm0.23)+(3.0\pm0.17)\log (GGN/r_{\rm GGN}),
\end{equation}
and
\begin{equation}
\log T_X=(-0.97\pm0.15)+(1.13\pm0.10)\log (GGN/r_{\rm GGN}),
\end{equation}
where $GGN/r_{\rm GGN}$ is in unit of Mpc$^{-1}$.

Most of the cluster richnesses are determined based on the galaxy count
\citep[e.g.,][]{aco89}. Other efforts were made to measure the richness
of cluster by various methods, e.g., correlation function amplitude of
the galaxies and the matched filter richness \citep{yl99,rrk+08}.
However, few measured richnesses tightly correlate with measurements in
X-rays due to the lack of accurate membership discrimination. With
membership discrimination using photometric redshift, we show
tighter correlations between the measurements of clusters in optical
and X-ray bands. Obviously, the accuracy membership discrimination is
crucial for finding the scaling relation of clusters.

\subsection{Correlations of the richness and optical luminosity with the cluster mass}

\begin{figure}
\epsscale{1.1}
\plotone{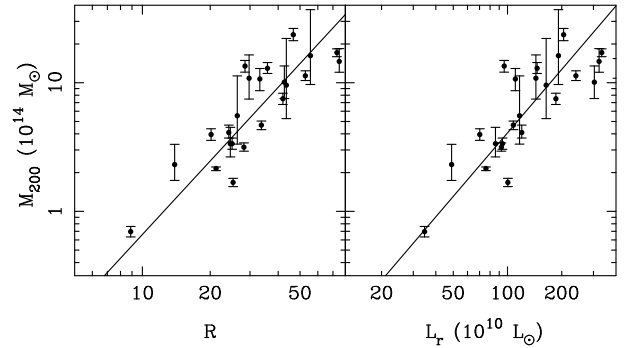}
\caption{Correlations between the cluster richness (the left panel) and
  summed $r$-band luminosity (the right panel) with cluster mass from
  \citet{rb02} for 24 clusters. The lines are the best fit given in
  Equations~(\ref{mr})--(\ref{ml})
\label{lum_ms}}
\end{figure}

The X-ray luminosity and temperature have been found to be tightly
correlated with cluster mass \citep{frb01,asf01,xjw01,rb02,sks+03}. As
shown, the cluster richness and summed luminosity are well correlated
with X-ray luminosity and temperature, hence can also trace cluster
mass.

We obtain cluster masses, $M_{200}$, determined from X-ray measurements
by \citet{rb02}. Here, $M_{200}$ is the total mass within the radius
$r_{200}$. Among the sample of \citet{rb02}, there are 28
clusters/groups in the sky region of SDSS DR6. We exclude four nearby
groups ($z\le0.006$), which contain only one luminous elliptical
galaxies. Therefore, we have masses of 24 clusters in the redshift
range $0.02<z<0.2$, of which 12 clusters are found in our
catalog. To determine the $M_{200}$--$L_r$ and $M_{200}$--$R$
relations precisely, we also calculate the richness and summed
$r$-band luminosities for the rest 12 clusters by our method.
We plot the richness and the summed luminosity against the cluster
masses for 24 clusters in Figure~\ref{lum_ms} and find the best fit as
\begin{equation}
\log\Big(\frac{M_{200}}{10^{14}~M_{\odot}}\Big)
=(-2.08\pm0.06)+(1.90\pm0.04)\log R,
\label{mr}
\end{equation}
and
\begin{equation}
\log\Big(\frac{M_{200}}{10^{14}~M_{\odot}}\Big)
=(-2.67\pm0.07)+(1.64\pm0.03)\log L_{r,10}.
\label{ml}
\end{equation}

In the previous studies, the mass--richness relation (the so called 
halo occupation distribution in some literature) is described in a
power law, $R\propto M^{\beta}$, and the factor $\beta$ is expected
less than 1 from the simulations \citep[e.g.,][]{whs01}. Our result,
$1/\beta=1.90\pm0.04$, i.e., $\beta=0.53\pm0.01$, is in agreement with
$\beta=0.55\pm0.04$ found by \citet{mh02}, but significantly smaller
than $\beta=0.70\pm0.04$ found by \citet{prg03} and
$\beta=0.92\pm0.03$ found by \citet{pbb+07}.

The correlation of cluster mass with the optical luminosity, i.e.,
mass-to-light ratio $M/L$, is also very interesting. \citet{amb+98}
investigated a fundamental plane in nearby rich Abell clusters and
suggested that the $M/L$ is not constant. In general, the $M/L$ is
also described by a power law, $M/L\propto L^{\tau}$, with $\tau$ in
the range 0.2--0.4 \citep{bc02,gmm+02,lms03,rgd+04,pbb+05}. Our
result, $\tau+1=1.64\pm0.03$, i.e., $\tau=0.64\pm0.03$, is larger
than the normal $\tau$ range, but smaller than $\tau=0.8$ found by
\citet{bsk+07}.

Recall that our cluster finding algorithm can detect 60\% clusters of
$N_{\rm in}=16$, which corresponds to a mean output richness $R=13.5$
according to Equation~(\ref{rinout}). The rate increases to 90\% for
clusters of $N_{\rm in}=20$ with a mean $R=16.7$. Using
Equation~(\ref{mr}), the output catalog is therefore 60\% complete for clusters
with a mass $M_{200}\sim1.2\times10^{14}~M_{\odot}$, and 90\% complete
for clusters with a mass $M_{200}\sim2\times10^{14}~M_{\odot}$.

\subsection{Candidates for New X-ray clusters}

The {\it ROSAT} All Sky Survey detects 18,806 bright sources
\citep{vab+99} and 105,924 faint sources \citep{vab+00} in the
0.1--2.4 keV band, of which 495 extended sources in the northern
hemisphere and 447 extended sources in the southern hemisphere have
been identified as clusters and AGNs or stars \citep{bvh+00,bsg+04}.

\begin{figure}
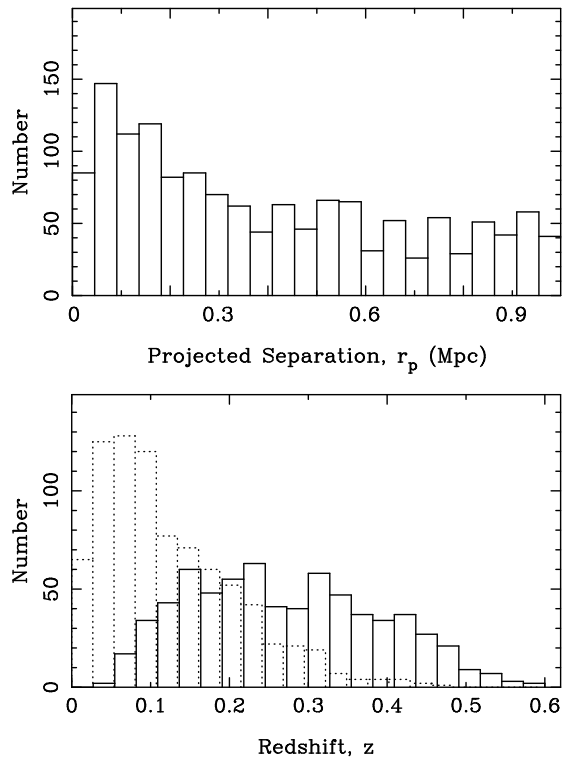

\epsscale{1.}
\plotone{f27a.eps}\\[2mm]
\plotone{f27b.eps}
\caption{Upper panel: Distribution of projected separations between
  {\it ROSAT} X-ray sources and the BCGs of clusters in our catalog. 
 Lower panel: Redshift distribution of the candidates for X-ray clusters
  ($r_p<0.3$ Mpc, solid histogram). The dotted histogram is for the
  known {\it ROSAT} X-ray clusters .
\label{histxc}}
\end{figure}

We cross-identify the {\it ROSAT} X-ray bright and faint sources with
clusters in our catalog to find new candidates for X-ray
clusters. Only those X-ray sources with a projected separation of
$r_p<0.3$ Mpc from the BCGs are probably associated with clusters (see
Figure~\ref{lxd}). The hardness ratios are expected in the range 0--1
for clusters \citep{bvh+00}, which can help to distinguish the cluster
X-ray sources. The X-ray sources with hardness ratios out of 0--1 can
be excluded to be associated with clusters. 
Figure~\ref{histxc} shows the distribution of a projected separation
between the X-ray sources and the BCGs of clusters in our catalog. If
the X-ray sources are uncorrelated with the clusters, the number of
pairs in each $r_p$ bin is proportional to $r_p^2$. The number excess
at low $r_p$ suggests that many of the X-ray sources are clusters. 

912 clusters in our catalog have a {\it ROSAT} X-ray source
within $r_p<0.3$ Mpc, and 227 of them are known X-ray clusters
according to NASA/IPAC Extragalactic Database. The rest 685 clusters
are new candidates for X-ray clusters. We notice that the candidate
distribution becomes constant within $0.5<r_p<1$ Mpc (see
Figure~\ref{histxc}). The real number of X-ray clusters should be the
excess over the constant level, them about 60\% of new candidates are
expected to be real X-ray clusters.  We also show the redshift
distribution of the new candidates for X-ray cluster together with
that of known {\it ROSAT} X-ray clusters from \citet{bvh+00,bsg+04} in
the lower panel of Figure~\ref{histxc}. Hundreds of candidates have
redshifts $z>0.3$. We list both 685 new candidates and 227 known X-ray
clusters in Table~\ref{xray} (a full list is available in the online
version).

\section{Summary}

We identify 39,668 clusters of galaxies in the redshift range
$0.05<z<0.6$ using photometric redshifts of galaxies from the SDSS
DR6. A cluster is recognized if more than eight member galaxies of
$M_r\le-21$ are found within a radius of 0.5 Mpc and a photometric
redshift gap between $z\pm0.04(1+z)$. This is the largest cluster
catalog to date. Our sample is much deeper in redshift than the
previous cluster catalogs from the SDSS. Cluster redshifts are
estimated with an uncertainty less than 0.022. Using the SDSS
spectroscopic data, we also estimate the contamination rate and
completeness of member galaxy candidates to be about 20\% and 90\%,
respectively. Monte Carlo simulations show that the cluster detection
rate depends on richness, but is approximately constant to redshift
$z=0.42$.  The detection rate is $\sim$60\% for clusters with a mass
$M_{200}\sim1.2\times10^{14}~M_{\odot}$, which corresponds to a mean
output richness $R\sim13.5$.  The detection rate increases to be 90\%
for clusters with a mass $M_{200}>2\times10^{14}~M_{\odot}$, which
corresponds to a mean $R\sim16.7$. The false detection rate of
clusters is $\sim$5\% for our algorithm.

We compare our catalog with the published Abell, CE, maxBCG, and {\it ROSAT}
X-ray cluster catalogs. We find that our catalog includes 77\% Abell
clusters and 77\% {\it ROSAT} X-ray selected clusters at $z>0.05$. Rich
clusters are more likely detected by our method.

With luminous member galaxies discriminated, we get the richness, $R$,
the summed luminosity, $L_r$, and the gross galaxy number $GGN$
within a cluster radius ($r_{\rm GGN}$) for clusters in our catalog up
to $z\sim0.42$. We find that they are tightly related to the X-ray
luminosity and temperature, and can trace the cluster mass with the
relations, $M_{200}\propto R^{1.90\pm0.04}$ and $M_{200}\propto
L_r^{1.64\pm0.03}$. By cross-identification with the {\it ROSAT} X-ray
source list, we obtain {\it 685 new candidates} of X-ray clusters, of which
60\% are likely true.

\acknowledgments

We thank the anonymous referee, Professor Y. Y. Zhou and Shude Mao for
valuable comments that helped to improve the paper.
The authors are supported by the National Natural Science Foundation
(NNSF) of China (10773016, 10821061 and 1083303) and the National Key
Basic Research Science Foundation of China (2007CB815403) and the
Doctoral Foundation of SYNU of China (054-55440105020).
Funding for the SDSS and SDSS-II has been provided by the Alfred
P. Sloan Foundation, the Participating Institutions, the National
Science Foundation, the U.S. Department of Energy, the National
Aeronautics and Space Administration, the Japanese Monbukagakusho, the
Max Planck Society, and the Higher Education Funding Council for
England. The SDSS Web site is http://www.sdss.org/.
The SDSS is managed by the Astrophysical Research Consortium for the
Participating Institutions. The Participating Institutions are the
American Museum of Natural History, Astrophysical Institute Potsdam,
University of Basel, Cambridge University, Case Western Reserve
University, University of Chicago, Drexel University, Fermilab, the
Institute for Advanced Study, the Japan Participation Group, Johns
Hopkins University, the Joint Institute for Nuclear Astrophysics, the
Kavli Institute for Particle Astrophysics and Cosmology, the Korean
Scientist Group, the Chinese Academy of Sciences (LAMOST), Los Alamos
National Laboratory, the Max Planck Institute for Astronomy (MPIA),
the Max Planck Institute for Astrophysics (MPA), New Mexico State
University, Ohio State University, University of Pittsburgh,
University of Portsmouth, Princeton University, the United States
Naval Observatory, and the University of Washington.

%\bibliographystyle{apj}
%\bibliography{journals,cluster}

\begin{deluxetable*}{rrrrrccrrrrrl}
%\tabletypesize{\tiny}
\tablecolumns{13}
\tablewidth{0pc}
\tablecaption{39668 Clusters identified from the SDSS DR6
\label{cat}}
\tablehead{ 
\colhead{Name}&\colhead{R.A.$_{\rm BCG}$}&\colhead{Decl.$_{\rm BCG}$} & \colhead{$z_p$}  &
\colhead{$z_{s,\rm BCG}$} & \colhead{$r_{\rm BCG}$} & \colhead{N$_{\rm gal}$} & \colhead{R}& $GGN$ &$r_{\rm GGN}$ 
&\colhead{$L_r$}    & \colhead{$D$} & \colhead{Other catalogs}\\
  &\colhead{(deg)}& \colhead{(deg)} & & & & & & & (Mpc) & \colhead{($10^{10}L_{\odot}$)} & &\\
\colhead{(1)} & \colhead{(2)} & \colhead{(3)} & \colhead{(4)} & 
\colhead{(5)} & \colhead{(6)} & \colhead{(7)} & \colhead{(8)} & 
\colhead{(9)} & \colhead{(10)}& \colhead{(11)}& \colhead{(12)}& \colhead{(13)}  \\
}
\startdata 
WHL J000006.0$+$152547& 0.02482&  15.42990& 0.1735&$-1.0000$& 16.58& 15& 11.30&  9.07& 0.50&  62.69& 6.97&  maxBCG\\                
WHL J000007.1$-$092909& 0.02957&$-9.48607$& 0.3963&$-1.0000$& 19.11& 19& 15.88& 14.44& 0.71&  81.33& 8.56&        \\                    
WHL J000007.6$+$155003& 0.03177&  15.83423& 0.1489&   0.1528& 16.00& 17& 13.40& 13.20& 0.71&  54.61& 6.55&  Abell,maxBCG\\            
WHL J000020.1$+$160859& 0.08358&  16.14976& 0.4591&$-1.0000$& 19.88& 20& 18.56& 29.40& 1.58& 107.01& 6.02&        \\                      
WHL J000021.7$+$150611& 0.09053&  15.10328& 0.2883&$-1.0000$& 17.67& 20& 18.17& 22.88& 1.50&  94.66& 9.43&  maxBCG\\                      
WHL J000025.1$-$093452& 0.10453&$-9.58125$& 0.3648&$-1.0000$& 18.44& 16&  9.29&  9.65& 0.71&  74.32& 4.90&        \\                      
WHL J000027.6$-$010140& 0.11617&$-1.04317$& 0.4491&   0.4387& 18.62& 25& 20.07& 20.07& 1.00& 124.10& 8.81&        \\                      
WHL J000048.3$-$011204& 0.18509&$-1.20016$& 0.4373&   0.4392& 18.76& 18& 14.44& 13.33& 0.87&  82.12& 5.01&        \\                      
WHL J000050.5$+$004705& 0.21051&   0.78477& 0.2458&$-1.0000$& 17.64& 22& 20.10& 26.16& 1.22& 105.53& 5.94&  NSCS,CE,maxBCG\\              
WHL J000050.7$+$004704& 0.21134&   0.78470& 0.4889&$-1.0000$& 19.73& 10&  6.40&  7.10& 0.50&  51.92& 5.69&        \\                      
WHL J000052.9$+$160520& 0.22045&  16.08902& 0.1986&$-1.0000$& 16.88& 12& 10.22& 11.33& 1.22&  46.42& 5.44&        \\                      
WHL J000059.1$+$004841& 0.24642&   0.81162& 0.3551&$-1.0000$& 19.18& 18& 14.80& 13.60& 0.87&  70.64& 4.93&  NSCS  \\                      
WHL J000111.3$+$151839& 0.29608&  15.30418& 0.4053&$-1.0000$& 19.10& 21& 19.10& 30.26& 1.58& 125.22& 6.50&        \\                      
WHL J000116.2$-$093137& 0.31767&$-9.52720$& 0.3383&   0.3693& 18.29& 24& 19.83& 31.65& 1.41& 112.56& 7.03&        \\                     
WHL J000117.5$+$142848& 0.32297&  14.48012& 0.3815&$-1.0000$& 19.68& 17& 12.84& 11.92& 0.71&  45.04& 4.93&        \\          
\enddata
\tablecomments{
Column (1): Cluster name with J2000 coordinates of cluster center; 
Column (2): R.A. (J2000) of cluster BCG; 
Column (3): Decl. (J2000) of cluster BCG; 
Column (4): photometric redshift of cluster; 
Column (5): spectroscopic redshift of cluster BCG, $-1.0000$ means not available; 
Column (6): $r$-band magnitude of cluster BCG; 
Column (7): number of member galaxy candidates within 1 Mpc;
Column (8): cluster richness;
Column (9): gross galaxy number;
Column (10): radius of member galaxy detection (Mpc);
Column (11): summed $r$-band luminosity of cluster;
Column (12): overdensity level of cluster;
Column (13): Other catalogs containing the cluster: 
Abell \citep{abe58,aco89}; Zwcl \citep{zhw68}; CE \citep{gsn+02}; NSC \citep{gcl+03}; 
NSCS \citep{ldg+04}; maxBCG \citep{kma+07b}; RXC \citep{bvh+00,bsg+04}.\\
{\it This table is available in its entirety in a machine-readable form in the online journal. 
A portion is shown here for guidance regarding the form and content.}
\\
{\bf The original catalog published in ApJS contains 39,716
clusters. Thanks Dr. Heinz Andernach for pointing out repeated entries 
(caused by a found bug in very early version of a code). We hence revised 
this table, and removed 48 entries and corrected BCG positions of 296 clusters.}
}
\end{deluxetable*}

\begin{deluxetable*}{ccrrcccl }
%\tabletypesize{\tiny}
\tablecolumns{8}
\tablewidth{0pc}
\tablecaption{912 clusters with ROAST X-ray sources including 685 candidates and 227 known X-ray clusters.
\label{xray}}
\tablehead{ 
\colhead{Name of X-ray source} & \colhead{Cluster Name}  & \colhead{R.A.$_{\rm BCG}$}  & \colhead{Decl.$_{\rm BCG}$} & 
\colhead{$z_p$} & \colhead{$r_p$} & \colhead{Count rate} & \colhead{Known X-ray clusters}       \\
                &              & \colhead{(deg)}   & \colhead{(deg)} &           & 
\colhead{(Mpc)} & \colhead{(count s$^{-1}$)} & \\
\colhead{(1)} & \colhead{(2)} & \colhead{(3)} & \colhead{(4)} & \colhead{(5)} & 
\colhead{(6)} & \colhead{(7)} & \colhead{(8)}  \\
} 
\startdata
RXS J000522.7$+$161306 &WHL J000524.0$+$161309 &   1.34987 &  16.21922 & 0.1115 &  0.04 &  0.076 & RXC \\    
RXS J001739.4$-$005150 &WHL J001740.0$-$005314 &   4.40670 &  $-0.87835$ & 0.2340 &  0.21 &  0.037 &     \\    
RXS J002302.3$+$144645 &WHL J002300.7$+$144656 &   5.75279 &  14.78240 & 0.3826 &  0.13 &  0.033 &     \\    
RXS J002815.2$+$135601 &WHL J002819.8$+$135459 &   7.08254 &  13.91657 & 0.1516 &  0.23 &  0.026 &     \\    
RXS J003209.2$-$003932 &WHL J003212.1$-$003950 &   8.04672 &  $-0.66670$ & 0.2175 &  0.14 &  0.013 &     \\    
RXS J003417.8$+$005145 &WHL J003419.1$+$004948 &   8.59684 &   0.85723 & 0.2035 &  0.27 &  0.024 &     \\    
RXS J004149.7$-$091817 &WHL J004148.2$-$091703 &  10.46029 &  $-9.30313$ & 0.0560 &  0.01 &  4.079 & RXC \\    
RXS J010101.1$-$095726 &WHL J010101.5$-$095717 &  15.25645 &  $-9.95473$ & 0.1457 &  0.03 &  0.035 &     \\    
RXS J010243.0$+$010805 &WHL J010243.1$+$010810 &  15.67950 &   1.13633 & 0.1345 &  0.01 &  0.052 & RX  \\    
RXS J010649.5$+$010317 &WHL J010650.5$+$010410 &  16.71051 &   1.06970 & 0.2527 &  0.21 &  0.187 & RXC \\    
RXS J010717.9$+$141635 &WHL J010721.9$+$141623 &  16.84109 &  14.27322 & 0.0963 &  0.10 &  0.020 &     \\    
RXS J010921.7$+$005457 &WHL J010923.1$+$005429 &  17.34616 &   0.90818 & 0.2723 &  0.14 &  0.029 &     \\    
RXS J011006.0$+$135849 &WHL J011001.3$+$135555 &  17.51321 &  13.97815 & 0.0712 &  0.06 &  0.061 & RXC \\    
RXS J011202.7$-$004355 &WHL J011204.1$-$004351 &  18.01689 &  $-0.73108$ & 0.2119 &  0.07 &  0.053 &     \\    
RXS J011940.0$+$145303 &WHL J011938.3$+$145352 &  19.90952 &  14.89799 & 0.1289 &  0.12 &  0.096 & RXC \\
\enddata
\tablecomments{
Column (1): Name of {\it ROSAT} X-ray source with J2000 coordinates; 
Column (2): Cluster name in Table~\ref{cat}; 
Column (3): R.A. (J2000) of cluster BCG; 
Column (4): Decl. (J2000) of cluster BCG; 
Column (5): photometric redshift of cluster; 
Column (6): projected separation between X-ray source and cluster BCG in Mpc; 
Column (7): Count rate of X-ray cluster in 0.1--2.4 keV band;
Column (8): Known X-ray clusters in the NASA/IPAC Extragalactic Database (NED).\\
{\it This table is available in its entirety in a machine-readable form in the online journal. 
A portion is shown here for guidance regarding the form and content.}
\\
{\bf The original catalog published in ApJS contains 790 candidates of
X-ray clusters, of which many known X-ray clusters in the NED were mixed in.
Thanks Dr. Heinz Andernach for pointing out this problem after publication. 
}
}
\end{deluxetable*}

\end{document}